\documentclass[twocolumn]{aastex631}

\usepackage{enumitem}
\usepackage{amsmath}
\usepackage{graphicx}
\usepackage{multirow}
\usepackage{stackengine}
\usepackage{url}
\usepackage{tablefootnote}
\usepackage[normalem]{ulem}
\newcommand\numberthis{\addtocounter{equation}{1}\tag{\theequation}}

\received{3/7/2025}
\revised{6/18/2025}
\accepted{7/1/2025}
\submitjournal{\apj}

\begin{document} 

\title{X-ray Measurements of $^{44}$Ti in Four Supernova Remnants}

\correspondingauthor{}
\email{thanove@clemson.edu}

\author[0009-0009-7793-9087]{Tyler E. Hanover}
\affiliation{Clemson University \\
Department of Physics and Astronomy \\
Clemson, SC 29634-0978, USA}

\author[0009-0008-2949-862X]{Mark D. Leising}
\affiliation{Clemson University \\
Department of Physics and Astronomy \\
Clemson, SC 29634-0978, USA}

\begin{abstract} 
We present the results of a search for scandium K$_\alpha$ X-ray fluorescence resulting from $^{44}$Ti electron capture decay in four young supernova remnants.
We analyzed archival \textit{XMM-Newton} data from Tycho's SNR, Kepler's SNR, Cassiopeia A, and SN 1987A. We fit the X-ray spectra in regions of interest with up to four components, thermal and power-law, as needed to fit the spectrum, and add a gaussian to test for extra Sc emission. While all require the line with modest confidence, none represent definitive detections on their own, and all are consistent with previous gamma-ray detections or limits on the $^{44}$Ti mass. 

\end{abstract}


\section{Introduction} \label{sec:intro}

In the effort to understand how stars explode as supernovae and what new elements they provide to the interstellar medium, radioactivity plays a key role.
Newly created unstable nuclei are unambiguous products of the explosion, can provide quantitative diagnostics of the explosive nucleosynthesis, and in some cases can clarify the location and dynamics of those nuclear burning regions.
Titanium-44 is a prime example of such a nucleus, with a decay lifetime that spans the transition from supernova to supernova remnant (SNR).
It results from the alpha-rich freeze-out of nuclear statistical equilibrium \citep{1973ApJS...26..231W} and is ejected among the last matter to reach escape velocity in core-collapse supernovae.
It might also be produced in explosive helium burning in thermonuclear (Type Ia) supernovae, especially if surface detonations trigger the central runaway. \citep{1986ApJ...301..601W,2020ApJ...888...80L}.

The classic Type Ia model begins with a binary system consisting of a carbon-oxygen white dwarf (WD) and a companion main-sequence or giant star that has overflowed its Roche lobe.
Material accretes onto the WD and causes its mass to grow.
When the WD nears the Chandrasekhar mass ($\approx$\,1.4\,M$_\odot$), internal pressure increases to the point of igniting carbon burning,
resulting in a thermonuclear runaway due to the inability of the electron-degenerate WD to expand and compensate for increased pressure from fusion-generated energy.

A core-collapse supernova, responsible for observational supernova Types Ib/Ic/II, results from the collapse of a massive star's iron core.
At lower core masses, {in the current paradigm,} the core undergoes rapid electron capture, the star explodes via a rebound shockwave fueled by neutrino flux and acoustic pressure waves, and a neutron star results \citep{RevModPhys.62.801}.

Supernova remnants are nebulae consisting of ejecta from the supernovae as well as circumstellar and interstellar material shocked by their blast waves. % and ISM
They radiate across nearly the entire electromagnetic spectrum creating radio synchrotron emission, optical emission line filaments, and conspicuous X-ray emission, and contain sufficiently long-lived radioactivity  produced in the explosions. 

$^{44}$Ti decays by electron capture \citep{1957PhRv..106...90H}
to $^{44}$Sc and then mainly by $\beta_+$ decay to $^{44}$Ca.
\begin{align*} 
    ^{44}\text{Ti} + e^- & \rightarrow\, ^{44}\text{Sc} + \nu_e \numberthis \\
    ^{44}\text{Sc} & \rightarrow\, ^{44}\text{Ca} + e^+ + \nu_e \numberthis
    \label{eqn:decay}
\end{align*}

Immediately after electron capture, the  $^{44}$Sc nuclei almost always de-excite by emitting a 78.4 keV gamma-ray followed by a 67.9 keV gamma-ray.
The electron capture leaves a K-shell vacancy, and the subsequent electronic de-excitation has a K$_\alpha$ fluorescent yield of 0.172 photons per decay.
The final decay to $^{44}$Ca, with positron emission (94\% probability), results in the emission of an 1157 keV gamma-ray.

$^{44}$Ti has been clearly detected in gamma-rays, as first predicted by \citet{1969ApJ...155...75C}, in two core-collapse SNRs: Cassiopeia A \citep{Iyudin94,Grefenstette17} and Supernova 1987A \citep{Grebenev12,Boggs2015}.
Such radioactive isotopes might also be detectable by measuring the fluorescence X-rays, though challenges are expected
\citep{2001ApJ...563..185L}.
It is the Sc K$_\alpha$ fluorescent line, whose energy depends on ionization state, from 4.09 keV (Ca-like) to 4.32 keV (He-like), for which we search in this study. % I believe 4.52 keV Sc Ly_alpha is unlikely
$^{44}$Ti has also possibly been detected in the youngest known galactic SNR, G1.9+0.3 \citep{Borkowski10} in the K$_\alpha$ line of scandium (however, see also \citet{2024MNRAS.529..999W}).

A survey with \textit{INTEGRAL} on the 68 \& 78 keV lines by \citet{Tsygankov16} constrains upper limits for young galactic SNRs in D. A. Green's catalogue\footnote{See \url{https://www.mrao.cam.ac.uk/surveys/snrs}} and
a study on the same lines by \citet{Weinberger20} attempts to constrain the progenitor types of many SNRs, including the four we look at in this work.
They both find that only Cassiopeia A has significant gamma-ray line flux and the other SNRs do not.

In this work, we use data from \textit{XMM-Newton} because of its large effective area in the 3--7 keV range compared to other X-ray observatories.
While, for example, \textit{Chandra}'s angular resolution is superior and \textit{NuSTAR}'s energy range permits direct measurement of some of the gamma-rays, the large effective area of \textit{XMM} makes it the best observatory for this work.

\section{Supernova Remnants of Interest} \label{sec:SNRs}

% relatively young, bright, and close SNRs
The half-life of $^{44}$Ti is $58.9\pm0.3$ yr \citep{Ti44-half-life}, so a key factor in searching for SNRs of interest is age.
The remnants need to be nearby and young enough for $^{44}$Ti to still be detectable. This constraint limits our search to remnants of supernovae within the past few centuries in the Milky Way and its satellites. 

Here we provide a brief summary of these SNRs and discuss why we chose to analyze them.
The ages and distances of these remnants can be found in Table \ref{table:SNRs}.

% Age as of 2024
\begin{deluxetable}{cccc}[hb]
    \tablehead{\colhead{Object} & \colhead{Age} & \colhead{Distance} & \colhead{Source for}\\[-2mm]
    \colhead{Name} & \colhead{(yr)\tablenotemark{a}} & \colhead{(kpc)} & \colhead{Distance}}
    \tablecaption{SNRs studied in this work\label{table:SNRs}}
    \startdata
    Tycho's SNR & 452 & $\approx2.5$ & \cite{Zhang13} \\ 
    Kepler's SNR & 420 & $4.8\pm1.4$ & \cite{Reynoso99} \\ 
    Cassiopeia A & 344 & $3.4_{-0.1}^{+0.3}$ &  \citet{Reed95} \\ 
    SN 1987A & 37.5 & $51.4\pm1.2$ & \cite{Panagia91} \\ 
    \enddata
    \tablenotetext{a}{Approximate date of supernova to September 2024}
\end{deluxetable}
\vspace{-18mm}

\subsection{Tycho's SNR} \label{subsec:tycho}
This remnant, hereafter Tycho, is
named after the Danish astronomer Tycho Brahe who observed the  supernova event in late 1572.
It is a well-studied historical object, and analysis of its light-echo spectrum by \citet{Krause08}  shows its progenitor was of Type Ia.
Our interest in Tycho is due to the possible presence of $^{44}$Ti in the remnant.
\citet{Wang14} report a flux excess in the 60-90 keV range using \textit{INTEGRAL}, and \citet{Troja14} obtain a similar result with \textit{Swift}/BAT.
However, \citet{Lopez15} do not detect  the 68 \& 78 keV lines with \textit{NuSTAR} at a flux level of $2.0\,\times\,10^{-5}\,\text{s}^{-1}\,\text{cm}^{-2}$. % assuming a distance of 2.3 kpc.

\subsection{Kepler's SNR} \label{subsec:Kepler}
Also named after its most prominent observer, Kepler's SNR (hereafter, Kepler) is the remnant of SN 1604.
Reconstruction of its light curve by \citet{Baade43} and evidence for the absence of a compact object \citep{ReynoldsS07} strongly support the notion that Kepler was also of Type Ia.
To our knowledge, no one has reported the detection of $^{44}$Ti in Kepler, but a detection is not impossible at the sensitivity we expect to achieve.

\subsection{Cassiopeia A} \label{subsec:cas-a}
Cas A is the remnant of a core-collapse SN \citep{Tananbaum99}, possibly observed by John Flamsteed in 1680 \citep{Ashworth80}.
It is the only galactic SNR with a clear detection of $^{44}$Ti, which makes it an obvious candidate for this work.
The first $^{44}$Ti-related detection was the 1157 keV line seen by \citet{Iyudin94} using \textit{COMPTEL}. % 4.1 sigma 
The 68 \& 78 keV lines were originally seen by \citet{Vink01} with \textit{BeppoSAX} % 3.4 sigma
and later confirmed by \citet{Renaud06} with \textit{INTEGRAL}. % 8.3 sigma
This emission was then spatially mapped by \citet{Grefenstette14} using \textit{NuSTAR}. % 4.9 sigma

\subsection{SN 1987A} \label{subsec:1987A}
The well-studied supernova of February 1987 in the Large Magellanic Cloud, first announced by \citet{Kunkel87} and shown to be a core collapse by many observations, including neutrino detections \citep{Arnett1989}, is a prime candidate for detection of radioisotopes because of its young age. 
Four lines from $^{56}$Co decay \citep{1988Natur.331..416M,1990ApJ...357..638L}, %\citep{Mahoney88},
the 122 keV line from $^{57}$Co \citep{Kurfess92},
and the 68 \& 78 keV lines from $^{44}$Ti \citep{Grebenev12, Boggs2015} have all been observed.
The two $^{44}$Ti gamma-ray lines were seen using both \textit{INTEGRAL} and \textit{NuSTAR}.

\subsection{G1.9+0.3} \label{subsec:g1.9}
This paper will focus on these four SNRs,
but it is still worth briefly discussing of G1.9+0.3 because the presence of $^{44}$Ti is unclear in this SNR.
First detected by \citet{Green84}, it is the youngest known galactic SNR at an age of $142\pm19$ years, assuming a distance near that of the galactic center of $\sim\,$8.5 kpc \citep{Luken20, ReynoldsS08}.
The Sc K$_\alpha$ line is reported by \citet{Borkowski10} with high significance using \textit{Chandra}.
However, neither \textit{NuSTAR} \citep{Zoglauer15} nor \textit{INTEGRAL} \citep{Tsygankov16} see the 68 \& 78 keV or 1157 keV gamma-rays, respectively.
We do not analyze G1.9+0.3 in this work as it has no dedicated \textit{XMM-Newton} observations.

\section{Data Analysis} \label{sec:Data}

Data were downloaded from NASA's HEASARC (\S\ref{ack}) and processed using \textit{XMM-Newton}'s Scientific Analysis System (SAS).
Data from the European Photon Imaging Camera (EPIC), specifically the pn-CCD, were analyzed.
Tables detailing the observations used can be found in Appendix \ref{append:obs}.

\subsection{XMM SAS} \label{subsec:SAS}
We began the EPIC data processing by filtering out high-count-rate background events.
Individual X-rays were detected as events by the CCD, and the SAS task \texttt{evselect} allowed us to filter the list of all events and then bin them by various parameters to create products such as light curves, images, and spectra.
Luminous background filtering was done by creating light curves for the 10--12 keV energy band and eliminating all events during times with significant rate excesses.
Light curves and filtered event files were created using \texttt{evselect} with flags \texttt{PATTERN==0} and \texttt{PATTERN$<$=4}, respectively.

Filtered event files were then binned into images to select those from particular regions of interest.
Images were created with \texttt{evselect} using no flags and a bin size of  {4 arcsec (\texttt{ximagebinsize$=$yimagebinsize$=$80})}.

Spectra were then created for each defined region of interest using the metatask \texttt{especget} (which runs \texttt{evselect}, \texttt{arfgen}, and \texttt{rmfgen}) with \texttt{extendedsource=yes} set and the flags \texttt{FLAG==0} and \texttt{PATTERN$<$=4} set.
The task \texttt{arfgen} generated the ancillary %(or auxiliary) 
response file that accounted for the effective area of the detector, and
\texttt{rmfgen} generated the redistribution (or response) matrix file that accounted for how the photon energies were distributed in measured count energies. 

All spectra collected from the same region were combined across observations with the task \texttt{epicspeccombine} for increased signal-to-noise, except in the case of SN 1987A, whose spectrum is evolving rapidly due to its young age.
Observation details for this SNR can be found in Appendix \ref{append:obs}, and hereafter we will refer to each one by its first three digits as (luckily) they are all unique.

\subsection{Spectral Regions} \label{subsec:regions}
There are many ways to subdivide a SNR, and each is likely to be interesting. 
Here we seek to detect $^{44}$Ti, so we analyze spectra from the full remnants because we do not necessarily know where to expect it.
We also select central regions with minimally bright X-ray emission from shocked gas because the bright continuum and thermal lines degrade our sensitivity to the weak Sc K$_\alpha$ line.
Of course, one cannot avoid the shocked gas in the foreground and background of the central ejecta. 

\begin{figure*}
    \centering
    \stackunder[5pt]{\includegraphics[width=0.24\textwidth]{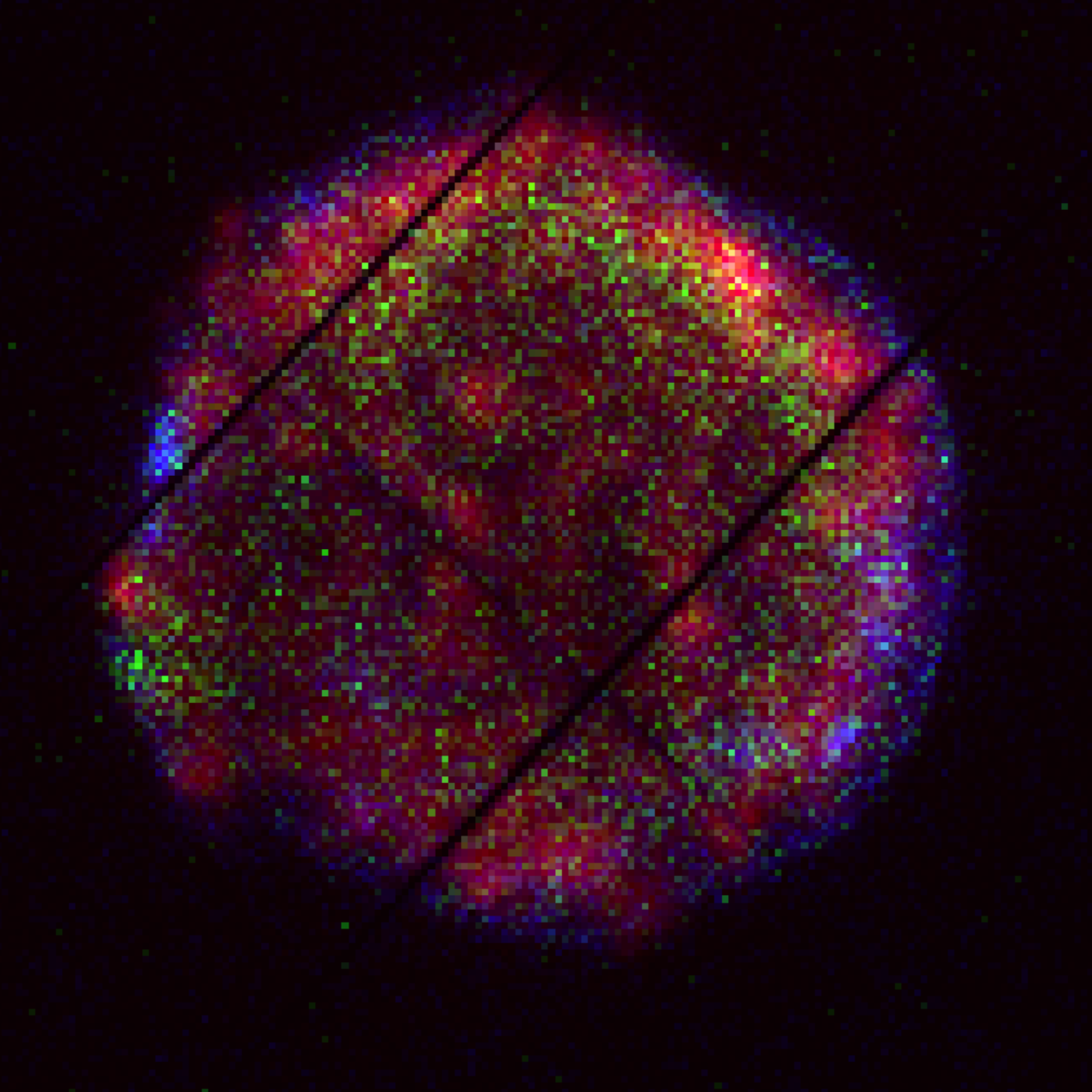}}{(a) Tycho -- Obs 0412380301}%
    \hfill%
    \stackunder[5pt]{\includegraphics[width=0.24\textwidth]{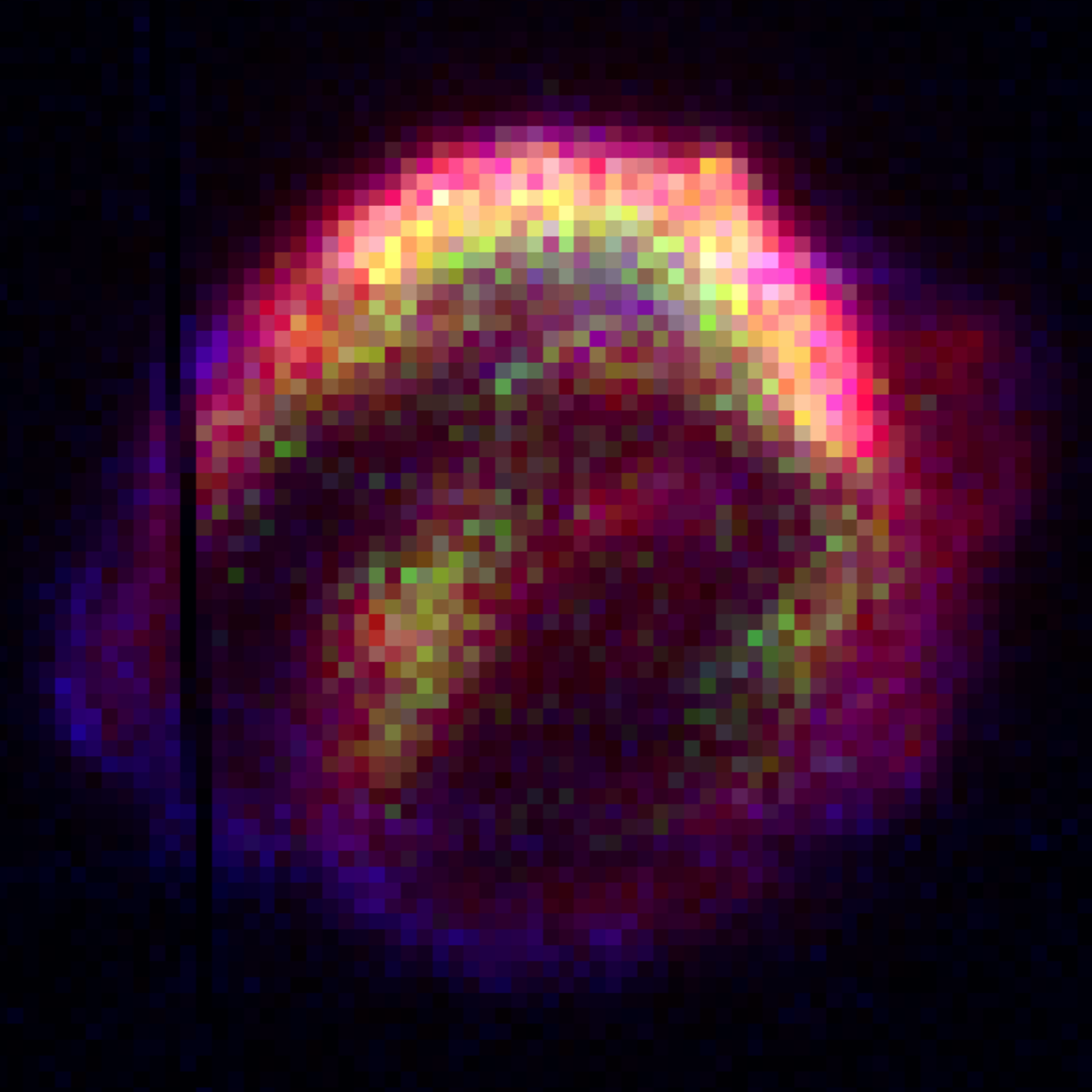}}{(b) Kepler -- Obs 0842550101}%
    \hfill%
    \stackunder[5pt]{\includegraphics[width=0.24\textwidth]{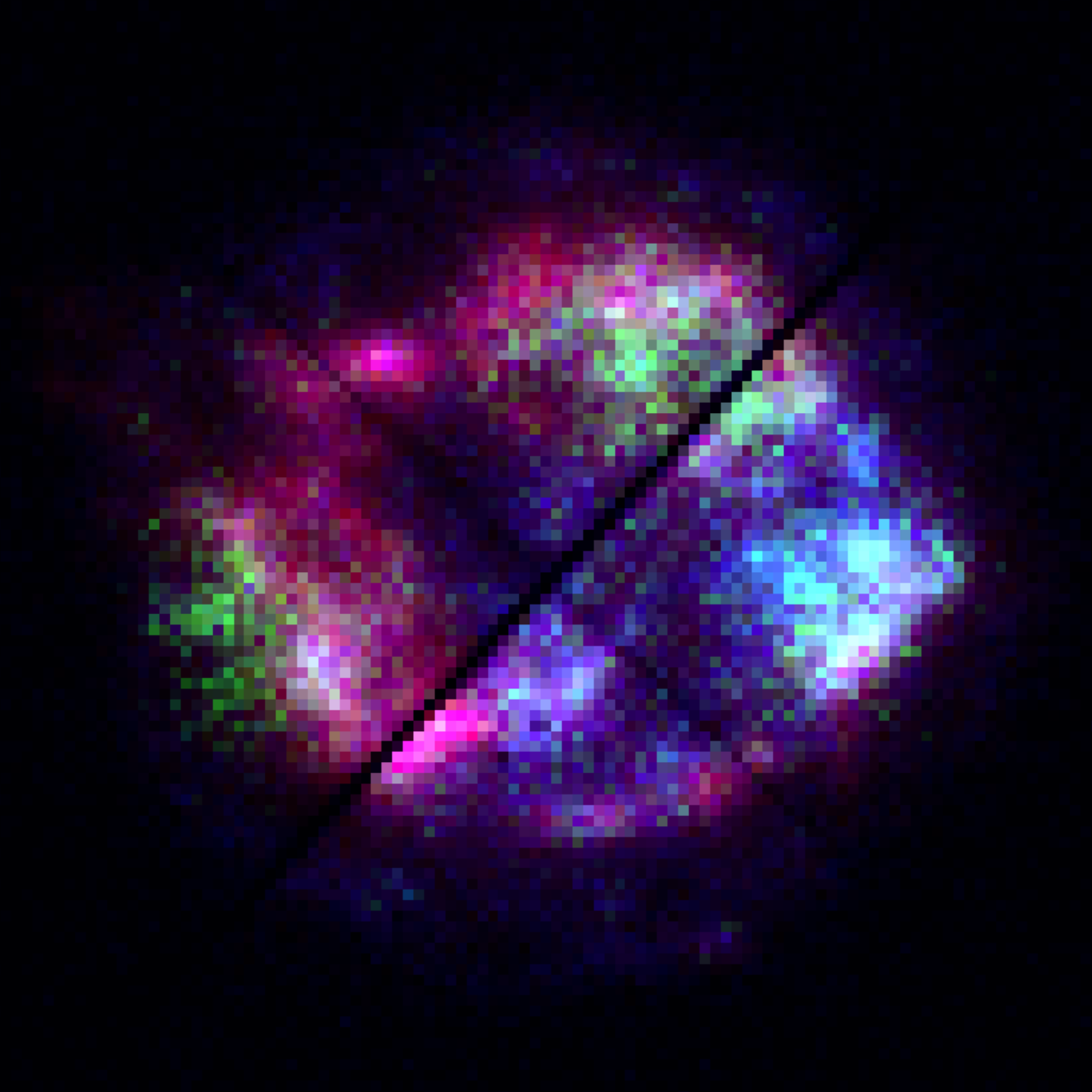}}{(c) Cas A -- Obs 0165560101}%
    \hfill%
    \stackunder[5pt]{\includegraphics[width=0.24\textwidth]{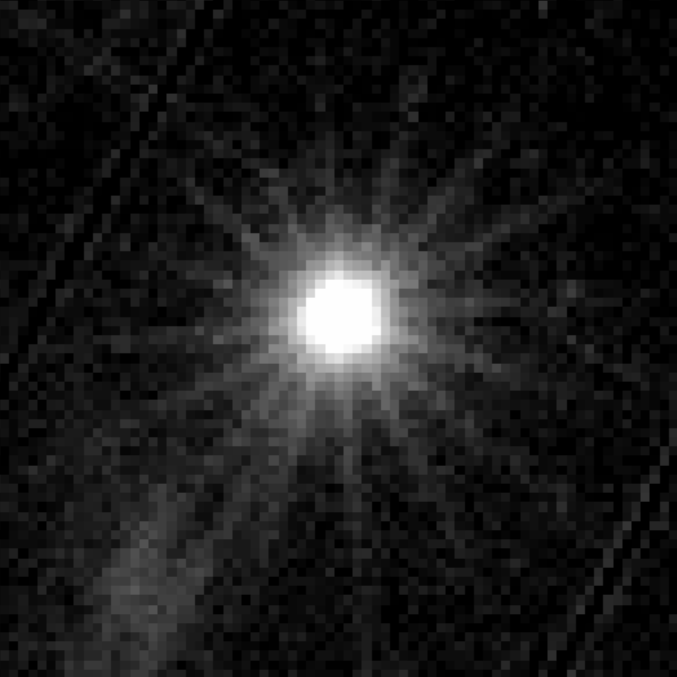}}{(d) SN 1987A -- Obs 090}
    \caption{EPIC-pn images of the four SNRs where red is 1.65--2.05 keV (Si K$_\alpha$ emission), green is 6.1--6.7 keV (Fe K$_\alpha$ emission), and blue is 4.1--6.1 keV (synchrotron-dominated continuum) for (a), (b), and (c).}%
    \label{fig:images}
\end{figure*}

\begin{figure*}
     \centering
    \stackunder[5pt]{\includegraphics[width=0.24\textwidth]{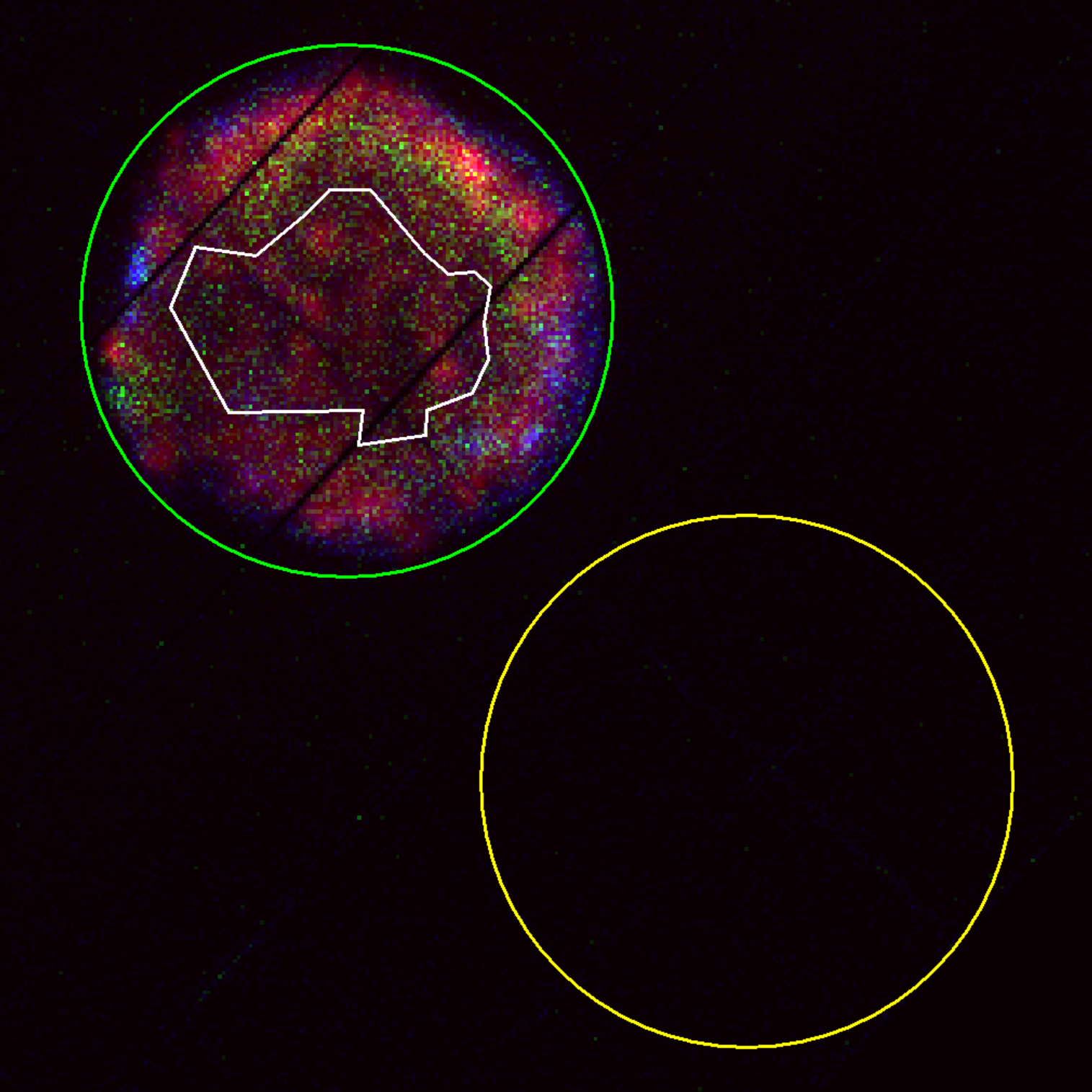}}{}%
    \hfill%
    \stackunder[5pt]{\includegraphics[width=0.24\textwidth]{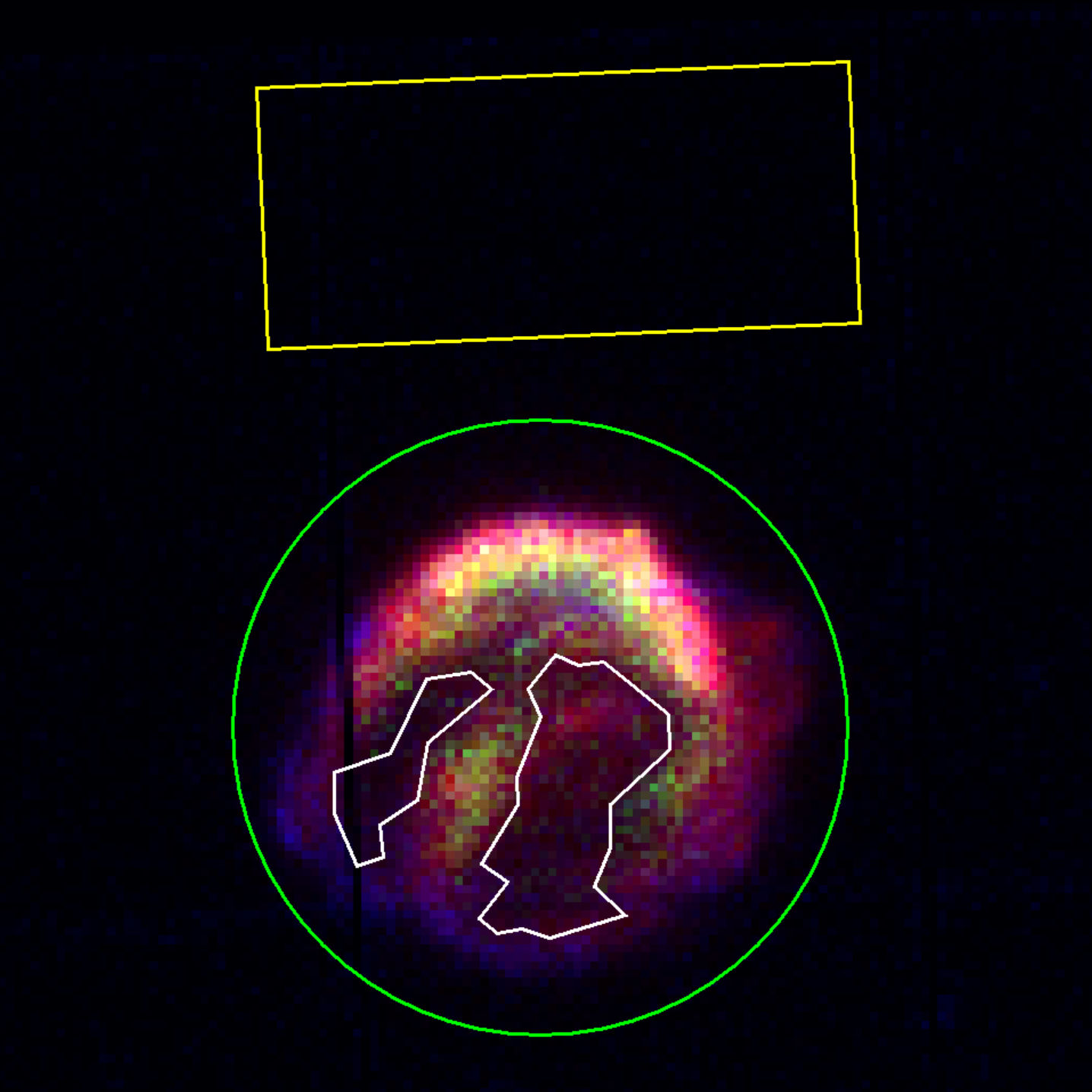}}{}%
    \hfill%
    \stackunder[5pt]{\includegraphics[width=0.24\textwidth]{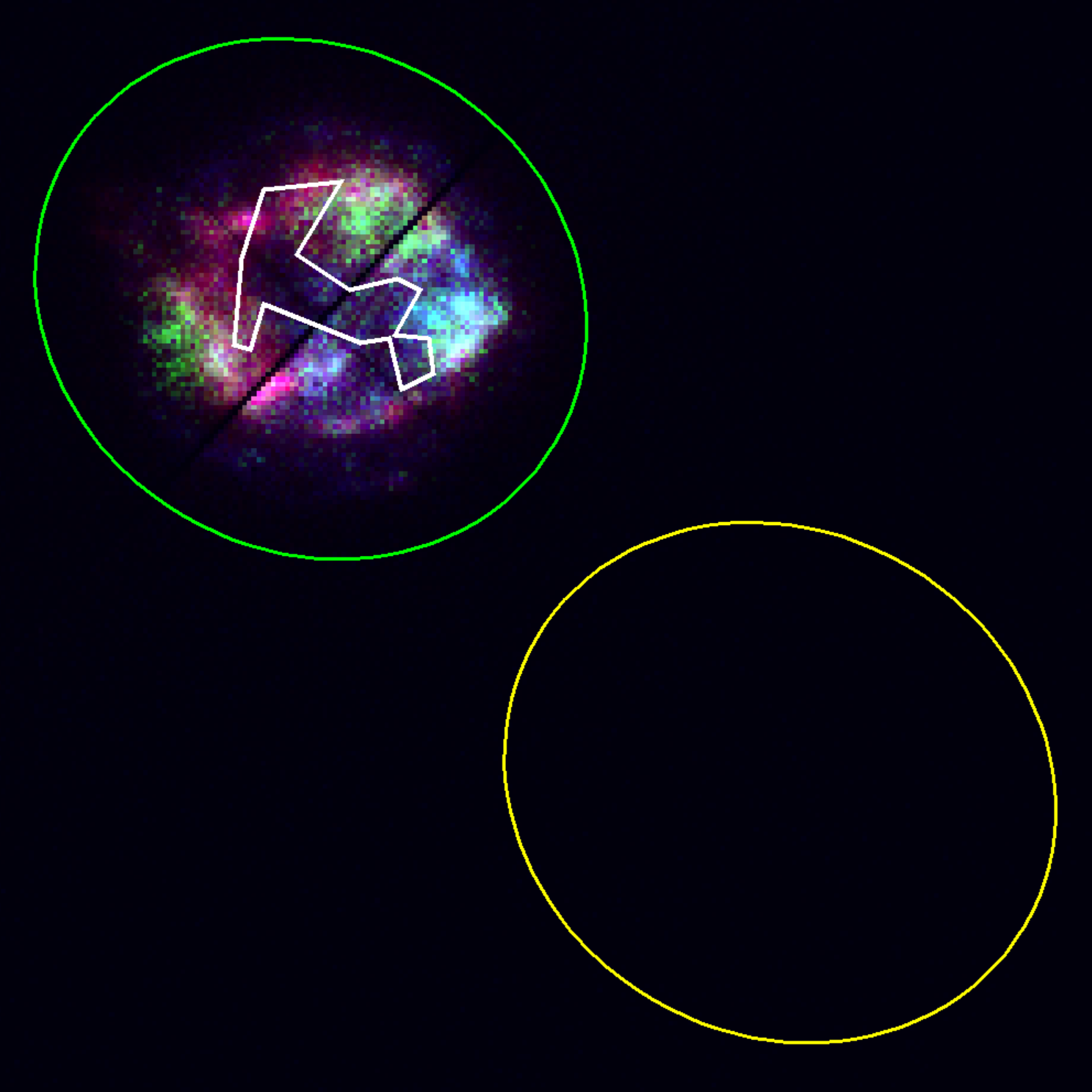}}{}%
    \hfill%
    \stackunder[5pt]{\includegraphics[width=0.24\textwidth]{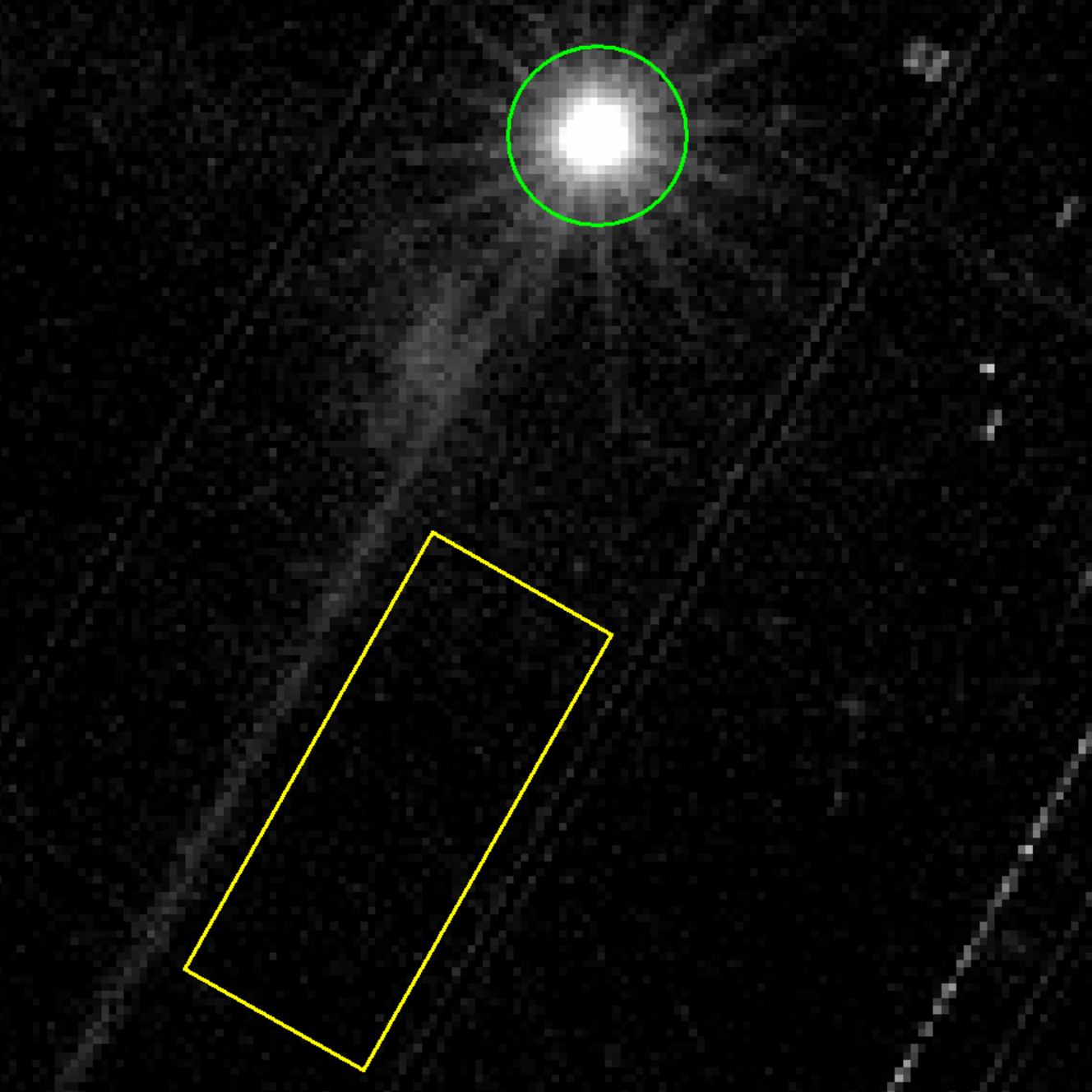}}{}
     \caption{Images from Figure \ref{fig:images} showing a wider area on the detector. The green regions are the full remnants, the white regions are the interiors, and the yellow regions are the backgrounds.}
     \label{fig:regions}
\end{figure*}

Images of each SNR are shown in Figure \ref{fig:images} and images with the selected regions are shown in Figure \ref{fig:regions}.
For SN 1987A, three-color images were not needed to see structure as the unresolved remnant is effectively a point source to {\it XMM}, and 
for Cas A, we attempt to closely match the shape of the spatially mapped $^{44}$Ti gamma-ray emission from \citet{Grefenstette14} as our interior region while avoiding the bright knot on the western side of the SNR.
Background spectra were taken from nearby regions, ideally on the same chip(s) as the source, with no noticeable source flux.

\section{Spectral Analysis} \label{sec:analysis}
All spectral fitting is done in XSPEC \citep{XSPEC}.
We model the spectra primarily with a combination of XSPEC's \textit{nei} and \textit{pshock} models \citep{Borkowski01}. %\footnote{See XSPEC's \textit{equil} model for other relevant references}.
Both are non-equilibrium ionization (NEI) collisional plasma models whose parameters include: plasma temperature $kT$, elemental abundances relative to solar ($Z\le30$), ionization timescale $\tau = n_e t$, redshift $z$, and a normalization
factor.\footnote{See \textit{nei} model in the XSPEC user manual for further details}
These two models are very similar except \textit{pshock} implements upper and lower limits for $\tau$ instead of a single value. In all of our fits, we set the lower limit for ${\tau}_{pshock} =$ $10^{8}\,\text{cm}^{-3}\,s$. % and the upper limit is allowed to vary.
In these models, we generally find that the \textit{pshock} components  correspond to plasmas with higher temperature and lower electron density while the \textit{nei} components are the reverse; however, this is not always the case.

We add a powerlaw model component to represent synchrotron emission and convolve the NEI models with the \textit{gsmooth} model to account for possible Doppler broadening.
In the tables below showing spectral fit parameters, ${\Delta \text{E}}_{\,6}$ represents the Doppler broadening (gaussian $1\sigma$ width) at 6 keV.
% The powerlaw's normalization factor is $}{ph.s^{-1}.keV^{-1}.cm^{-2}}$ at 1 keV.
In most cases, interstellar absorption is negligible as we fit the spectra only above 3 keV; however, in the cases where it is not, photoelectric absorption model \textit{phabs} is used.
 {Although absorption in the Milky Way remnants is negligible for our purposes, here we provide the average H I column densities from the HEASARC: $7.2\,\times\,10^{21}\,\text{cm}^{-2}$ in the direction of Tycho, $2.1\,\times\,10^{21}\,\text{cm}^{-2}$ in the direction of Kepler, and $2.7\,\times\,10^{21}\,\text{cm}^{-2}$ in the direction of Cas A.}
Finally, after finding the best fit for this multi-component model,  {an instrument-width} gaussian is added to represent the emission from Sc K$_\alpha$ and the model is refit.
The gaussian centroid is bounded between line energies 4.09 and 4.52 keV (soft upper limit of 4.32 keV).
 {The remnants themselves have bulk motion which may slightly change this line centroid, but we account for that in the uncertainties in the tables below.}
For a spectral fit, an example of our model is $gsmooth*vvpshock\, +\, gsmooth*vvnei\, +\, powerlaw\, +\, gaussian$ (a $v$ or $vv$ prefix allows for the elemental abundances to individually vary).

We allow for collisionally ionized stable Sc by setting its abundance relative to solar equal to that derived for Ti in all fits.
Stable Sc isotopes are likely produced in supernovae but in such small quantities that they are undetectable here --- the solar Sc abundance is four orders of magnitude lower than that of Fe \citep{AndersGrev89}, and so much lower than that of Ti, that even when a Ti line is visible in the spectrum, the corresponding stable Sc line is not in our current sample of spectra.
It is possible the inclusion of collisionally ionized Sc will mask a portion of the Sc line produced after $^{44}$Ti electron capture, but again the flux of this thermal feature should be negligible.
Setting the Sc abundance to 0 instead of tying it to Ti results in a negligible change to the fit statistic in our model.
We argue that a significant Sc K$_\alpha$ line would only be seen in a SNR spectrum if it is a direct result of $^{44}$Ti decay.

For this photon-counting detector, we use the C-Statistic (C-Stat), which was derived from  {likelihood ratios} by \citet{Cash79}, to test models attempting to describe the accumulated spectra.
% photon emission is random, chi^2 is for gaussian data but photons produce poisson data
All uncertainties for our model fits were calculated using XSPEC's  {Goodman-Weare MCMC method \citep{GoodmanWeare}} with 200 walkers, 2 million steps, and a burn-in of 1 million steps (in a few cases these values were doubled).
 {The actual number of steps in the Goodman-Weare algorithm is $(\text{total steps})/(\text{total walkers})$; therefore in our fits, each walker is taking 10,000 steps. We use such a large number of walkers because a vast majority of our fits have $>\,$20 free parameters.} % tycho 37 pars, kepler 21 pars, cas 28 pars, 87A had 13 or 22 pars
All confidence ranges and error-bars are 90\% unless stated otherwise.

We \textbf{do not} expect the elemental abundances in our model outputs to be the true abundances in these SNRs; the goal of this work is to use physically motivated thermal models that fit young SNR spectra well enough that we can search for a possible excess near the expected position of the Sc K$_\alpha$ line.

\subsection{Tycho's SNR Spectra} \label{subsec:tycho_spec}
\renewcommand{\arraystretch}{1.2}
\begin{table}[b]
    \centering
    \caption{Best fit for Tycho (full SNR)}
    \vspace{-2mm}
    \begin{tabular}{|c|c|c|c|}
        \hline
        Parameter & vvpshock & vvnei 1 & vvnei 2 \\
        \hline
        ${\Delta \text{E}}_{\,6}$\,(eV) & $50.5_{-4.7}^{+4.1}$ & $99.8_{-19.5}^{+14.3}$ & $29.5_{-15.5}^{+19.4}$ \\ 
        kT (keV) & $3.40_{-0.29}^{+0.26}$ & $2.26_{-0.26}^{+0.35}$ & $1.66_{-0.16}^{+0.22}$ \\
        Ar {\,/\,Ar$_\odot$} & $17.2_{-9.1}^{+9.9}$ & $116_{-34}^{+35}$ & $40.2_{-7.9}^{+10.4}$ \\
        Ca {\,/\,Ca$_\odot$} & $14.6_{-7.6}^{+10.3}$ & $148_{-48}^{+51}$ & $56.3_{-12.5}^{+17.0}$ \\
        Sc {\,/\,Sc$_\odot$} & = Ti {\,/\,Ti$_\odot$} & = Ti {\,/\,Ti$_\odot$} & = Ti {\,/\,Ti$_\odot$} \\
        Ti {\,/\,Ti$_\odot$} & $86.4_{-26.1}^{+26.7}$ & $\le80.2$ & $34.6_{-23.8}^{+29.8}$ \\ 
        Cr {\,/\,Cr$_\odot$} & $51.8_{-9.7}^{+13.8}$ & $140_{-62}^{+56}$ & $20.5_{-11.9}^{+12.9}$ \\ 
        Mn {\,/\,Mn$_\odot$} & $41.1_{-13.4}^{+14.6}$ & $\le90.1$ & $90.8_{-63.5}^{+73.7}$ \\ 
        Fe {\,/\,Fe$_\odot$} & $10.9_{-2.0}^{+2.3}$ & $20.4_{-12.1}^{+12.0}$ & $4.1_{-1.5}^{+1.5}$ \\
        $\tau$ $(10^{10}\ \text{cm}^{-3}\,\text{s})$ & $0.86_{-0.12}^{+0.17}$ & $2.38_{-0.50}^{+0.51}$ & $4.70_{-0.91}^{+1.18}$ \\ 
        z $(10^{-3})$ & $-2.71_{-0.34}^{+0.45}$ & $6.06_{-1.22}^{+0.94}$ & $-5.44_{-0.74}^{+1.47}$ \\
        norm $(10^{-3})$ & $30.3_{-5.0}^{+5.7}$ & $5.70_{-1.37}^{+1.48}$ & $13.9_{-3.1}^{+4.0}$ \\
        \hline
        \multicolumn{2}{|c|}{$\alpha_\text{ {powerlaw}}$} & \multicolumn{2}{|c|}{$2.53_{-0.07}^{+0.07}$} \\
        \multicolumn{2}{|c|}{$\text{norm}_\text{ {powerlaw}}\,(10^{-3})$} & \multicolumn{2}{|c|}{$38.8_{-5.9}^{+6.3}$} \\
        \hline
        \multicolumn{2}{|c|}{$\text{E}_\text{\,Gauss}$\,(keV)} & \multicolumn{2}{|c|}{$4.27_{-0.10}^{+0.04}$} \\
        \multicolumn{2}{|c|}{$\text{F}_\text{Gauss}$\,($10^{-6}\,\text{s}^{-1}\,\text{cm}^{-2}$)} & \multicolumn{2}{|c|}{$6.5_{-4.1}^{+2.9}$} \\
        \hline
        \multicolumn{4}{|c|}{C-Stat/dof = 870/802 (1.08)} \\
        \hline
    \end{tabular}
    \label{table:tycho_full}
\end{table}

\begin{figure}[ht]
    \centering
    \includegraphics[width=\columnwidth]{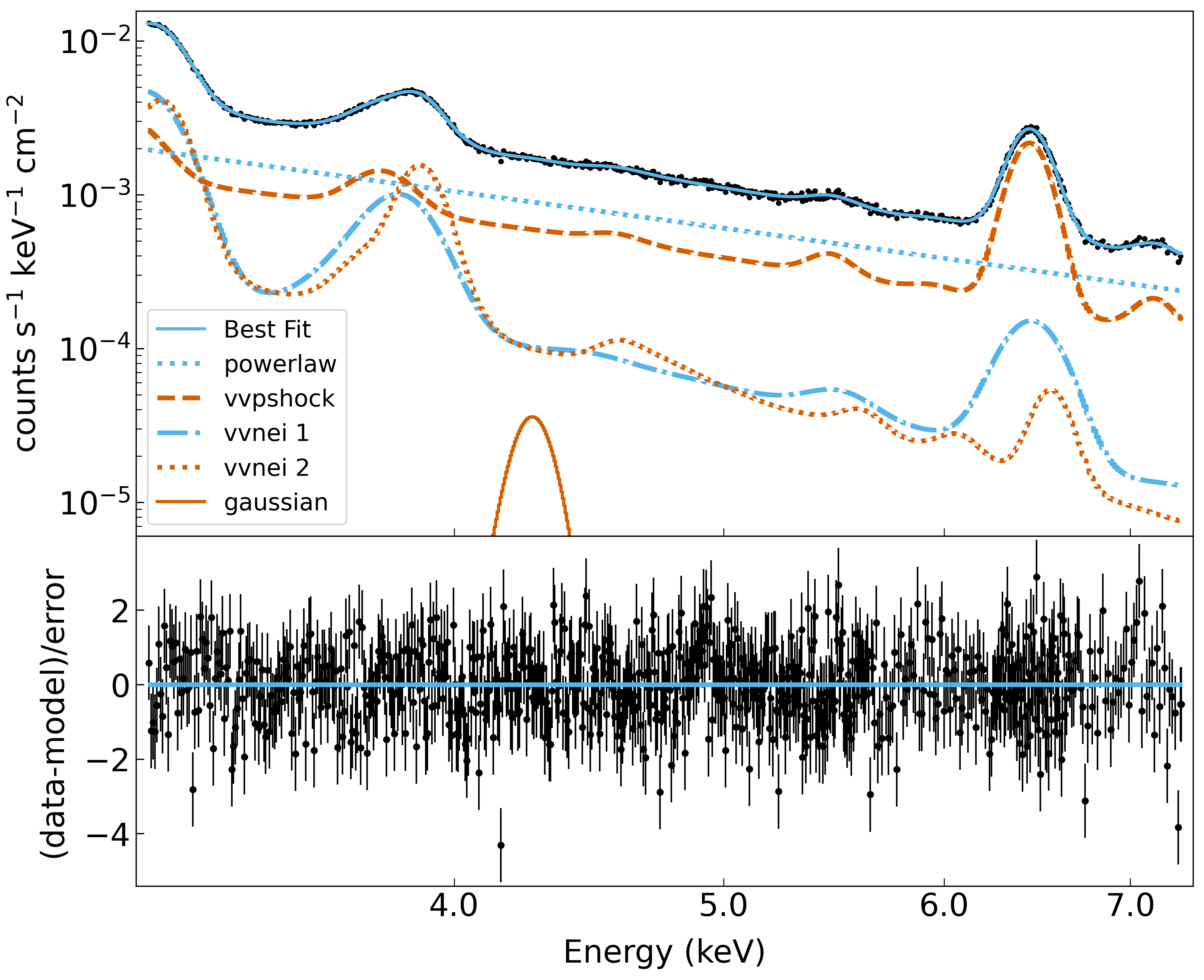}
    \includegraphics[width=\columnwidth]{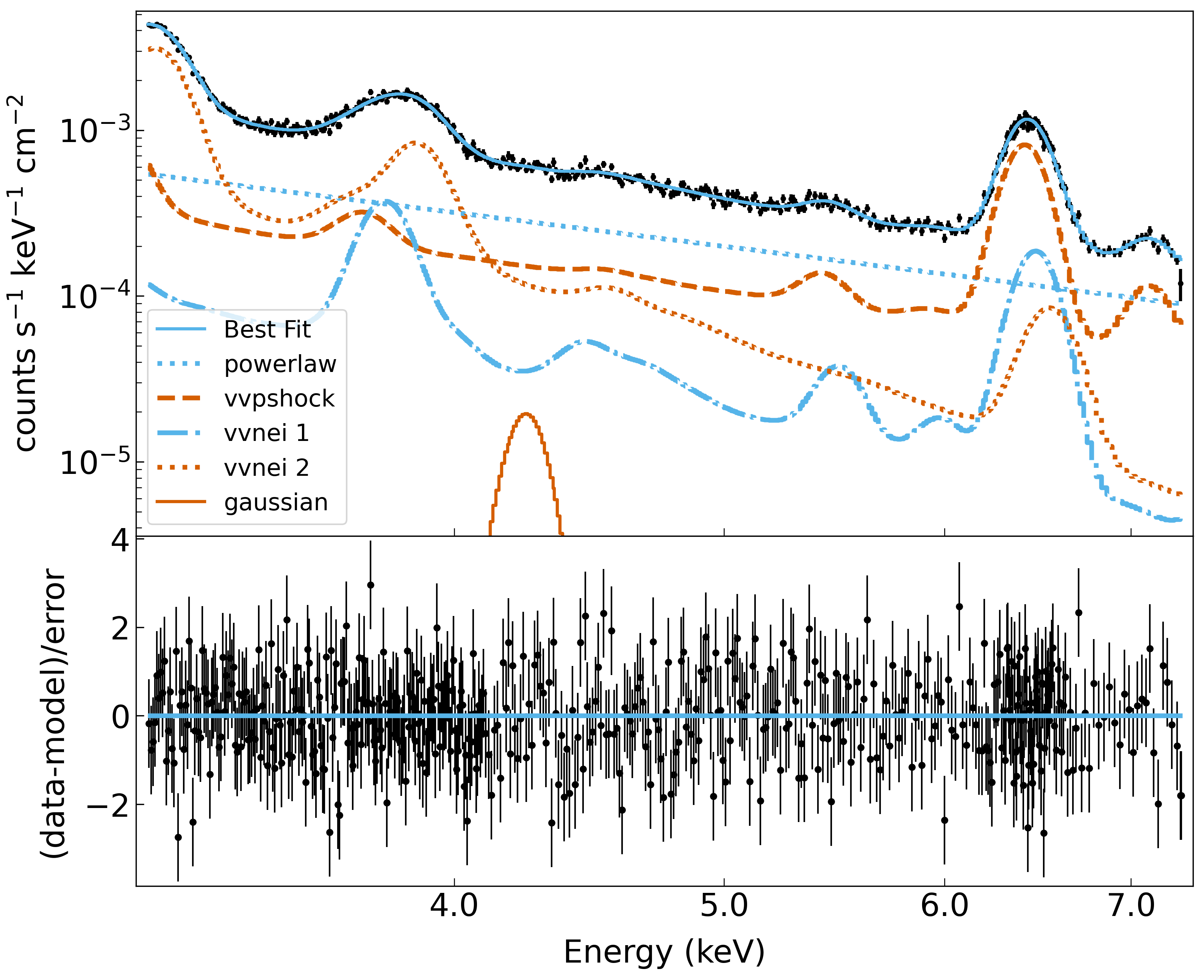}
    \vspace{-5mm} \caption{Spectral fits for Tycho's SNR from the Ar K$_\alpha$ line to the Fe K$_\beta$ line. The top plot is the full remnant and the bottom plot is the interior only.} %corresponding to Table \ref{table:tycho_full}}
    \label{fig:tycho_specs}
\end{figure}

\begin{figure}[ht]
    \centering
    \includegraphics[width=\columnwidth]{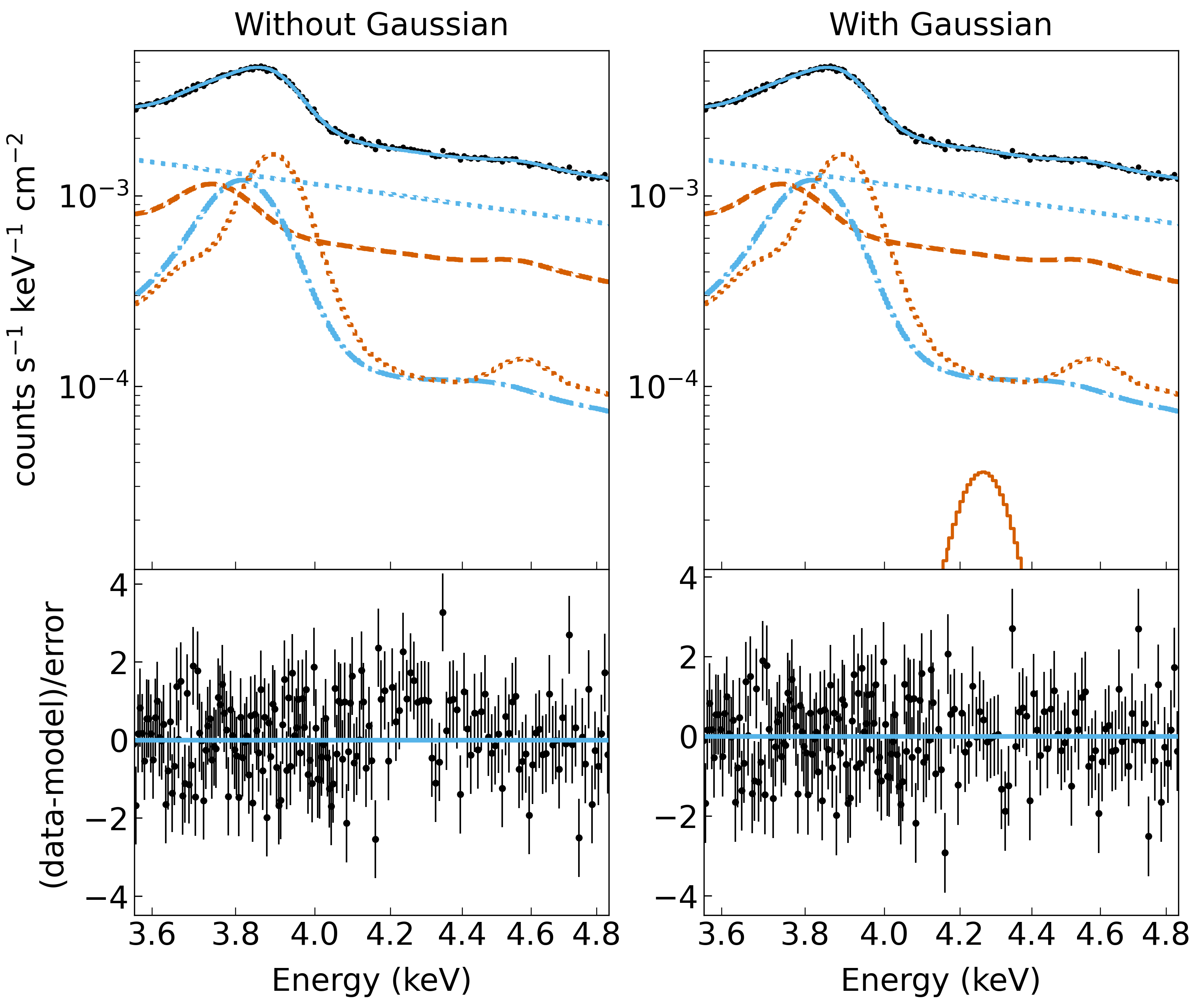}
    \caption{A portion of the spectral fit from the top plot in Figure \ref{fig:tycho_specs} shown with (right) and without (left) the Sc K$_\alpha$ gaussian included in the fit. One can see in the residuals on the left why the fit prefers a nonzero Sc line.} 
    \label{fig:gauss-compare}
\end{figure}

Twelve observations of Tycho, when merged, have a total exposure time of 258.4 ks after  filtering out background flares.
We begin fitting the spectra with a single NEI model that, even when tested in smaller spacial regions, unsurprisingly results in a poor fit because Tycho is a young SNR with varying temperatures and densities along most lines of sight.
The next logical step is to add a second NEI component  {(which can be \textit{pshock} or \textit{nei})} of a different temperature to account for the missing photons in the single-component model.
In many cases, a third NEI component is also needed to well represent the range of temperature and ionization states apparently present in this remnant.
We stop adding model components when the spectrum is reasonably well fit. One way to check for a reasonable fit is using XSPEC's \textit{goodness} command. For example, we simulate 100 spectra with \textit{goodness} from our best-fit model of the full Tycho SNR and it finds 74\% of those simulated spectra get a similar or lower C-Stat.
Adding more components is not useful for our purposes.
The spectral fits are shown in Figure \ref{fig:tycho_specs} and the corresponding best-fit parameters for the full SNR can be found in Table \ref{table:tycho_full}.
A comparison of the full remnant fit with and without the gaussian can be found in Figure \ref{fig:gauss-compare}.
An interesting result with the full remnant fit is that the three NEI components each contribute a significant amount of flux to the Ar K$_\beta$ -- Ca K$_\alpha$ feature, whereas the Fe features are more dominated by the $pshock$ component, which is the highest temperature of the three.

Our interior region in Figure \ref{fig:regions} is inspired by region 5 in \citet{Miceli2015} who found indications for the presence of shocked titanium in the outer regions of Tycho. 
They postulate that $^{48}$Ti and $^{50}$Ti are the predominant isotopes in the Ti emission line and that those lines may not coincide spatially with $^{44}$Ti.
We do note that their best-fit line energy of $\approx\,$4.9 keV would imply a strong blueshift as the line energy for He-like Ti is 4.75 keV (it would be unlikely to find a large abundance of H-like Ti even with the high temperatures in the remnant).
This region has weak continuum and Fe and Ca lines, presumably because the reverse shock has not yet heated that interior.
With low thermal emission to hide the Sc line there, this is a promising region to search for $^{44}$Ti decay if indeed $^{44}$Ti was produced and remains there.

\subsection{Kepler's SNR Spectra} \label{subsec:kepler_spec}
\begin{figure}[ht]
    \centering
    \includegraphics[width=\columnwidth]{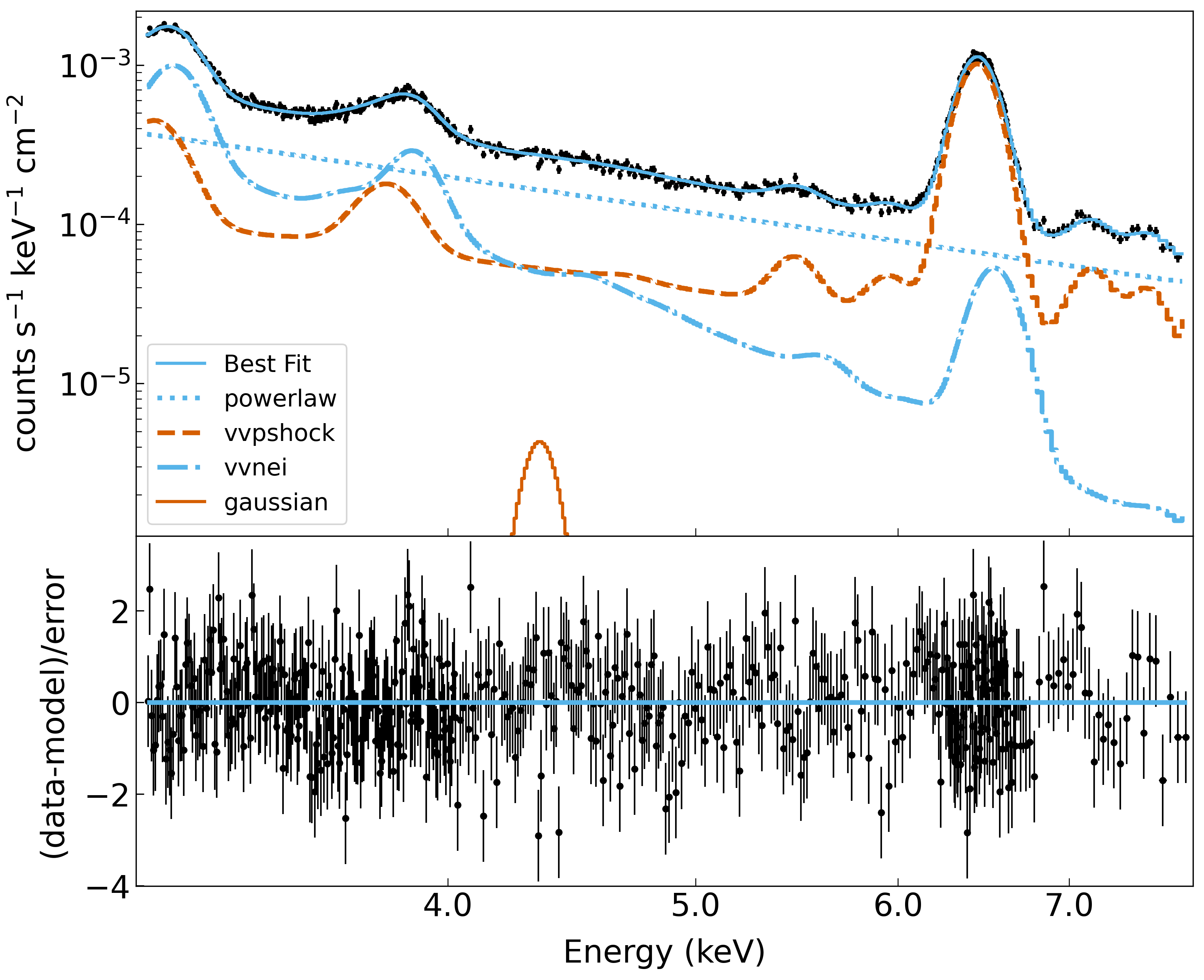}
    \includegraphics[width=\columnwidth]{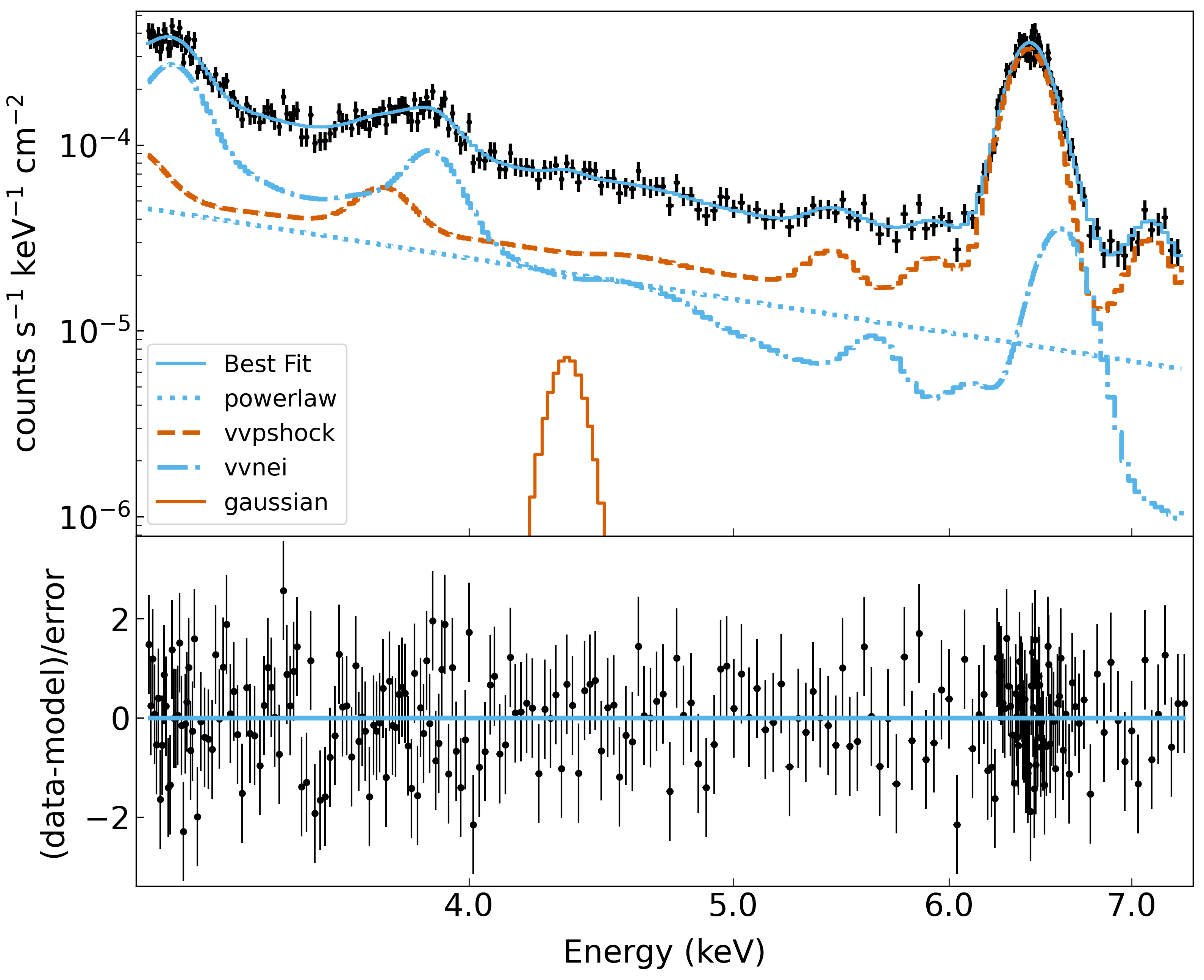}
    \vspace{-5mm} \caption{Spectral fits of Kepler's SNR from the Ar K$_\alpha$ line to the Ni K$_\alpha$ line. The top plot is the full remnant and the bottom plot is the interior only.} 
    \label{fig:kepler_specs}
\end{figure}

Three observations, when merged, have a total exposure time of 155.8 ks after background filtering on Kepler's SNR.
The spectral fits are shown in Figure \ref{fig:kepler_specs} and the best-fit parameters for the full remnant can be found in Table \ref{table:kepler_full}.
Kepler has the only spectrum  of those studied in this paper that clearly shows a resolved Ni K$_\alpha$ line. Cas A's nickel line is present but not easily distinguishable from its Fe K$_\beta$ line in the EPIC-pn spectra. Nickel lines in Tycho and 1987A are not seen clearly in their respective spectra, hence  the abundance is fixed at solar for those two SNRs.

Two thermal components were necessary to fit the spectrum but resulted in a C-Stat/dof less than 1, so we tie together the abundance values between components (e.g. Mn$_{nei}$ = Mn$_{pshock}$) to prevent further overfitting  {(allowing all the abundances to vary individually results in larger uncertainties and a C-stat/dof $<$ 0.9)}.
Tying abundances together for the other SNRs generally resulted in a poor fit due to their larger counts compared to Kepler. % counts = s^-1 keV^-1 cm^-2

We find Sc K$_\alpha$ $1\sigma$ flux upper limits of $2.1\,\times\,10^{-6}$ and $1.7\,\times\,10^{-6}\,\text{s}^{-1}\,\text{cm}^{-2}$ for the entire remnant and interior, respectively.
The gaussian line energy ranged from 4.11--4.50 keV because there is no clear excess in the spectrum.
This finding is also evident in the 90\% flux lower bounds being negative.

\begin{table}[ht]
    \centering
    \caption{Best fit for Kepler (full SNR)}
    \vspace{-2mm}
    \begin{tabular}{|c|c|c|c|}
        \hline
        Parameter & vvpshock & vvnei & powerlaw \\
        \hline
        ${\Delta \text{E}}_{\,6}$\,(eV) & $73.0_{-3.9}^{+2.7}$ & $84.4_{-8.9}^{+10.2}$ & $-$ \\ 
        kT (keV) & $5.32_{-0.90}^{+1.14}$ & $0.89_{-0.09}^{+0.16}$ & $-$ \\
        Ar {\,/\,Ar$_\odot$} & $5.0_{-1.0}^{+1.3}$ & = pshock & $-$ \\
        Ca {\,/\,Ca$_\odot$} & $5.8_{-1.3}^{+1.6}$ & = pshock & $-$ \\
        Sc {\,/\,Sc$_\odot$} & = Ti {\,/\,Ti$_\odot$} & = pshock & $-$ \\
        Ti {\,/\,Ti$_\odot$} & $22.1_{-20.0}^{+73.3}$ & = pshock & $-$ \\ 
        Cr {\,/\,Cr$_\odot$} & $17.6_{-4.7}^{+5.9}$ & = pshock & $-$ \\ 
        Mn {\,/\,Mn$_\odot$} & $25.1_{-7.7}^{+10.1}$ & = pshock & $-$ \\ 
        Fe {\,/\,Fe$_\odot$} & $6.6_{-1.6}^{+1.8}$ & = pshock & $-$ \\
        Ni {\,/\,Ni$_\odot$} & $7.4_{-2.4}^{+2.7}$ & = pshock & $-$ \\
        $\tau$ $(10^{10}\ \text{cm}^{-3}\,\text{s})$ & $1.59_{-0.37}^{+0.31}$ & $31.7_{-16.0}^{+35.0}$ & $-$ \\ 
        z $(10^{-3})$ & $-1.98_{-0.73}^{+1.51}$ & $0.24_{-2.29}^{+0.88}$ & $-$ \\
        $\alpha$ & $-$ & $-$ & $2.31_{-0.16}^{+0.19}$ \\
        norm $(10^{-3})$ & $1.27_{-0.30}^{+0.52}$ & $4.98_{-1.34}^{+1.33}$ & $4.31_{-1.39}^{+1.67}$ \\
        \hline
        \multicolumn{2}{|c|}{$\text{E}_\text{\,Gauss}$\,(keV)} & \multicolumn{2}{|c|}{$4.36_{-0.25}^{+0.14}$} \\
        \multicolumn{2}{|c|}{$\text{F}_\text{Gauss}$\,($10^{-6}\,\text{s}^{-1}\,\text{cm}^{-2}$)} & \multicolumn{2}{|c|}{$0.7_{-1.2}^{+2.1}$} \\
        \hline
        \multicolumn{4}{|c|}{C-Stat/dof = 888/915 (0.97)} \\
        \hline
    \end{tabular}
    \label{table:kepler_full}
\end{table}

\begin{figure}[hb]
    \centering
    \includegraphics[width=\columnwidth]{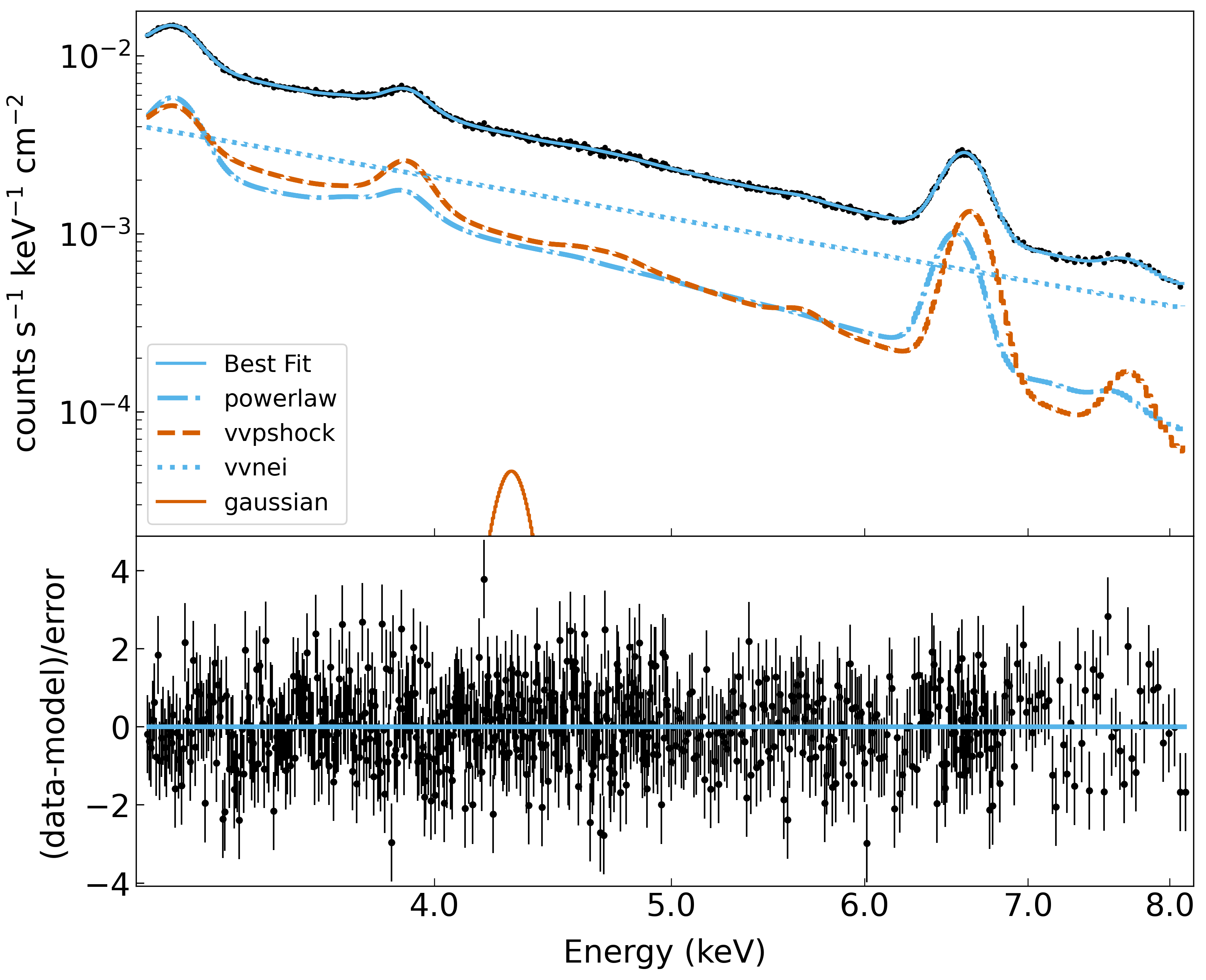}
    \caption{Spectral fit, from the Ar K$_\alpha$ line to the Fe K$_\beta$ line, for the spacial region corresponding to the NuSTAR-detected $^{44}$Ti emission in Cas A.} % corresponding to Table \ref{table:cas_a_nustar}}
    \label{fig:casA_spec}
\end{figure}

\subsection{Cas A Spectra} \label{subsec:cas_a_spec}
Cas A's twelve observations, when merged, have a total exposure time of 284.3 ks after background filtering.
The main region analyzed was that corresponding to the NuSTAR-detected $^{44}$Ti emission spatially mapped by \citet{Grefenstette14}.
For this spacial region, the spectral fit is shown in Figure \ref{fig:casA_spec} and the best-fit parameters are in Table \ref{table:cas_a_nustar}.
Adding a third NEI component to this fit did not improve the fit enough to justify keeping it in the model; even so, it is interesting that the two components are somewhat similar, whereas in other SNRs, the components have stark differences.
We find a gaussian line energy of $4.29_{-0.05}^{+0.06}\ $keV that would imply Li-like or He-like $^{44}$Sc.
This is not unreasonable as the best fit \textit{kT} for both NEI components implies temperatures of over 20 million Kelvin.

\begin{table}[ht]
    \centering
    \caption{Best fit for Cas A (NuSTAR's $^{44}$Ti Region)}
    \vspace{-2mm}
    \begin{tabular}{|c|c|c|c|}
        \hline
        Parameter & vvpshock & vvnei & powerlaw \\
        \hline
        ${\Delta \text{E}}_{\,6}$\,(eV) & $57.3_{-9.4}^{+10.9}$ & $65.7_{-5.3}^{+6.9}$ & $-$ \\ 
        kT (keV) & $2.23_{-0.25}^{+0.40}$ & $1.78_{-0.13}^{+0.15}$  & $-$ \\
        Ar {\,/\,Ar$_\odot$} & $5.0_{-2.9}^{+3.3}$ & $2.4_{-2.1}^{+1.8}$ & $-$ \\
        Ca {\,/\,Ca$_\odot$} & $2.9_{-2.4}^{+2.7}$ & $3.5_{-1.5}^{+1.4}$ & $-$ \\
        Sc {\,/\,Sc$_\odot$} & = Ti {\,/\,Ti$_\odot$} & = Ti {\,/\,Ti$_\odot$} & $-$ \\
        Ti {\,/\,Ti$_\odot$} & $\le21.2$ & $12.0_{-4.4}^{+6.0}$ & $-$ \\ 
        Cr {\,/\,Cr$_\odot$} & $1.1_{-1.0}^{+2.7}$ & $6.5_{-2.5}^{+3.0}$ & $-$ \\ 
        Mn {\,/\,Mn$_\odot$} & $\le3.8$ & $\le1.9$ & $-$ \\ 
        Fe {\,/\,Fe$_\odot$} & $4.2_{-1.2}^{+1.7}$ & $4.0_{-1.1}^{+1.4}$ & $-$ \\
        Ni {\,/\,Ni$_\odot$} & $5.8_{-5.1}^{+6.5}$ & $41.5_{-13.7}^{+19.5}$ & $-$ \\
        $\tau$ $(10^{10}\ \text{cm}^{-3}\,\text{s})$ & $6.10_{-2.06}^{+1.38}$ & $13.8_{-3.08}^{+4.04}$ & $-$ \\
        z $(10^{-3})$ & $-4.05_{-1.82}^{+1.16}$ & $-2.62_{-0.57}^{+0.78}$ & $-$ \\
        $\alpha$ & $-$ & $-$ & $2.44_{-0.08}^{+0.07}$ \\
        norm ($10^{-2}$) & $8.06_{-2.10}^{+2.52}$ & $10.7_{-2.6}^{+2.3}$ & $6.28_{-0.87}^{+1.23}$ \\
        \hline
        \multicolumn{2}{|c|}{$\text{E}_\text{\,Gauss}$\,(keV)} & \multicolumn{2}{|c|}{$4.29_{-0.05}^{+0.06}$} \\
        \multicolumn{2}{|c|}{$\text{F}_\text{Gauss}$\,($10^{-6}\,\text{s}^{-1}\,\text{cm}^{-2}$)} & \multicolumn{2}{|c|}{$12.0_{-6.6}^{+5.1}$} \\
        \hline
        \multicolumn{4}{|c|}{C-Stat/dof = 1131/1010 (1.12)} \\
        \hline
    \end{tabular}
    \label{table:cas_a_nustar}
\end{table}

\subsection{1987A Spectra} \label{subsec:1987A_spec}
\begin{figure}[ht]
    \centering
    \includegraphics[width=\columnwidth]{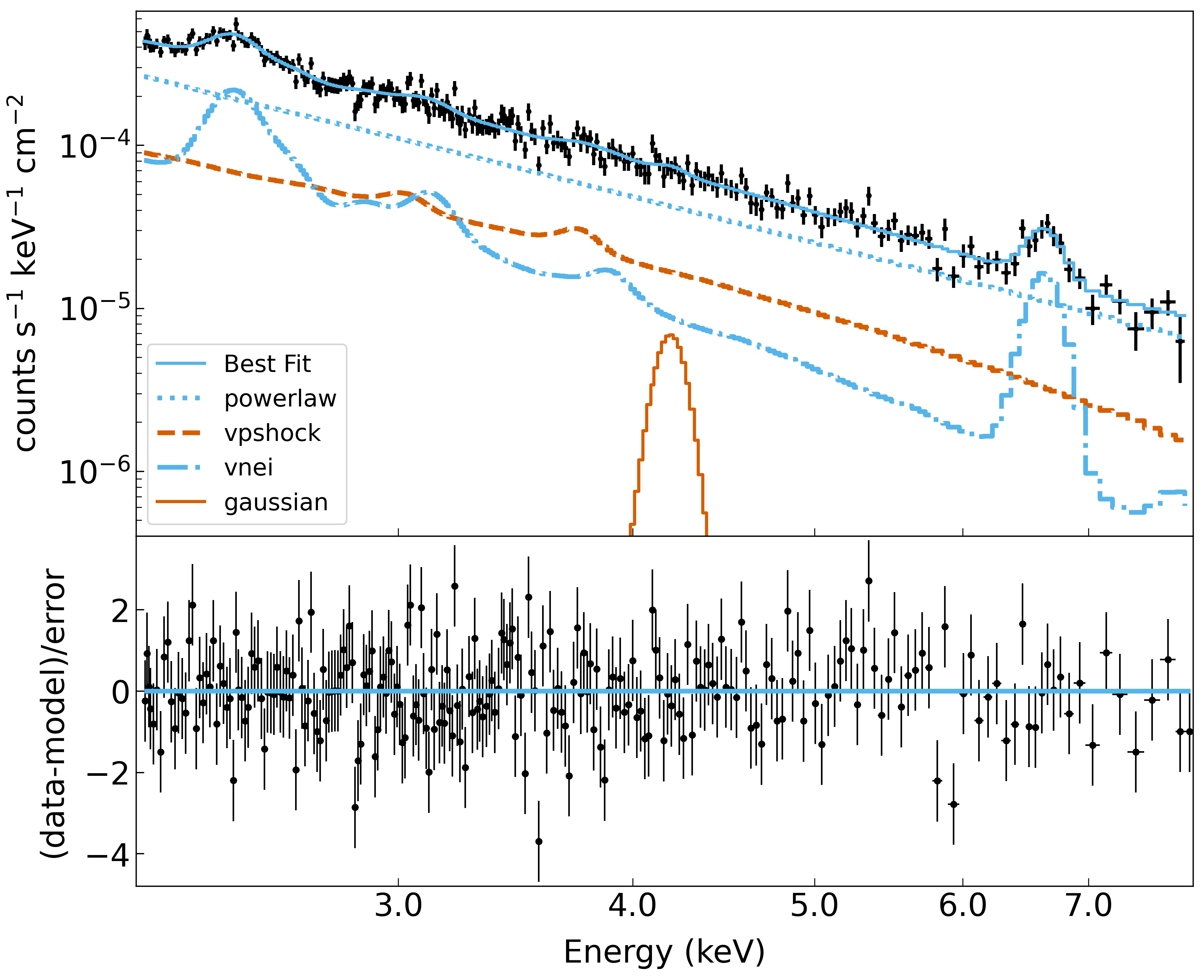}
    \caption{Spectral fit for SN 1987A observation 086 from the S K$_\alpha$ line to the Fe K$_\alpha$ line.}
    \label{fig:086_fit}
\end{figure}

\begin{figure}[hb]
    \centering
    \includegraphics[width=\columnwidth]{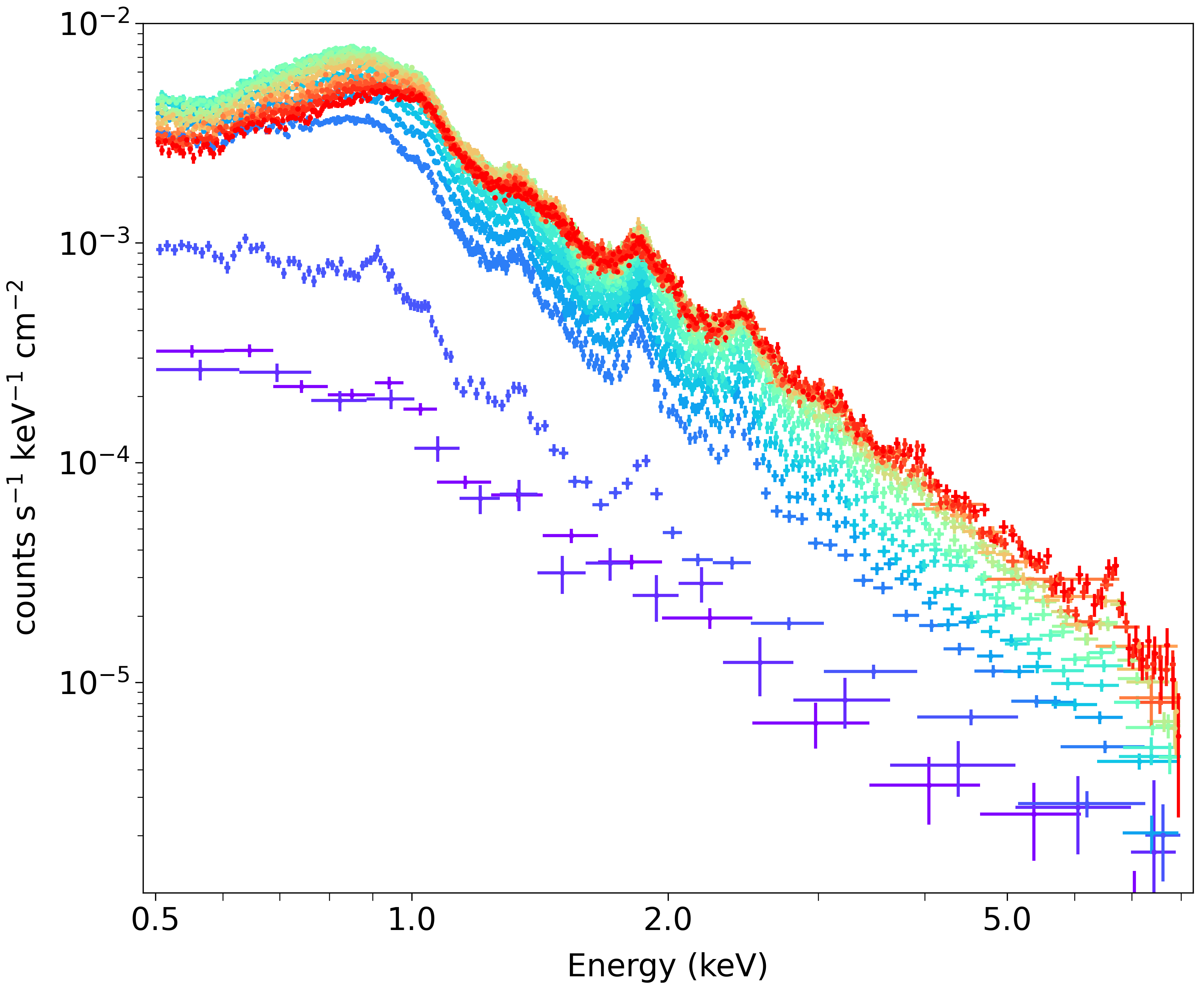}
    \caption{EPIC-pn spectra for the nineteen SN 1987A observations. The color gradient is earliest (blue/purple) to most recent (red).}
    \label{fig:all19}
\end{figure}

Nineteen observations for SN 1987A were analyzed; the average exposure time after background filtering is 47.8 ks.
These observations are not merged as the young age of this SNR causes the  flux and spectrum to change significantly over the time of these observations.
For these spectra, a 2--8 keV energy range was needed for a robust fit (the spectra are heavily continuum-dominated from 3--6 keV), making interstellar absorption non-negligible, so the entire model is multiplied by XSPEC's \textit{phabs} with n$_\text{H}$ fixed at $2.6\,\times\,10^{21}\,\text{cm}^{-2}$ \citep{Alp21}.
One of the spectral fits is shown in Figure \ref{fig:086_fit}, and all the spectra, plotted from 0.5--8.0 keV, are shown in Figure \ref{fig:all19}. Interestingly, flux in the band 0.5--1.1 keV begins to decrease after 2014.
Fit parameters for all of the SN 1987A fits can be found in Appendix \ref{append:87A}.
Only the single ``$v$'' prefix model version was used for this remnant rather than the double ``$vv$'' because there are no discernible Cr or Mn lines.

Each spectrum was initially fit using one NEI component, with most requiring a second component to achieve a reasonably good fit.
Abundances varied greatly because of low signal-to-noise in the thermal lines, especially in the early observations, but all the fits were good enough to add the gaussian and estimate a Sc K$_\alpha$ flux. The gaussian in Figure \ref{fig:086_fit} is not significant despite its apparent height, (($1.4_{-1.2}^{+0.7})\,\times\,10^{-6}\,\text{s}^{-1}\,\text{cm}^{-2}$).

\section{Results} \label{sec:results}
% Discuss findings
% Convert fluxes to masses
The main goal of this work is estimating the initial $^{44}$Ti mass produced in the four supernovae. Given the fluxes determined above, we apply
the flux-to-mass conversion  {estimate} for radioisotopes \citep[e.g.,][]{Lopez15} to the Sc K$_\alpha$ line: 
\begin{equation}
    M_{44} = 4\pi d^2 \,(44\,m_p) \,\tau \exp{(t/\tau)} \,F_{\text{Gauss}} \,W^{-1}
\end{equation}
where $d$ = distance to the source, $m_p$ = proton mass, $\tau$ = mean lifetime of the radioisotope, $t$ = age of the source, and $W$ = emission efficiency (mean number of photons per decay).

The half-life of $^{44}$Ti is $58.9\pm0.3$ yr, therefore its mean lifetime is $\approx85.0$ yr. %\pm0.43
The emission efficiency of the Sc K$_\alpha$ line after $^{44}$Ti electron capture decay is 0.172 photons/decay.
The results of these calculations can be found in Tables \ref{table:MW_flux} and \ref{table:87A_flux}.
 {Where appropriate, we} report negative flux lower limits as $0\,\text{M}_\odot$.
For the Milky Way SNRs, we use the age at the time of the most recent observation --- these are listed in Appendix \ref{append:obs}.

\renewcommand{\arraystretch}{1.2}
\begin{table*}
    \normalsize
    \centering
    \begin{tabular}{ccCCCC}
    
        \hline \hline
        
        \multirow{2}{*}{Object} & \multirow{2}{*}{Region} & \multicolumn{2}{c}{Line Flux ($10^{-6}\,\text{s}^{-1}\,\text{cm}^{-2}$)} & \multicolumn{2}{c}{Implied Mass ($10^{-4}\,\text{M}_\odot$)} \\[-1mm]
        & & 1\sigma & 2\sigma & 1\sigma & 2\sigma \\
        \hline
        
        \multirow{2}{*}{Tycho's SNR} & Entire Remnant & 6.5_{-2.5}^{+1.6} & 6.5_{-5.2}^{+3.6} & 5.2_{-2.0}^{+1.3} & 5.2_{-4.2}^{+3.0} \\ 
        & Interior Only & 3.4_{-1.8}^{+2.8} &  {\le 8.5} & 2.8_{-1.4}^{+2.3} & \le 6.9 \\ 
        \hline
        
        \multirow{2}{*}{Kepler's SNR} & Entire Remnant & 0.7_{-0.5}^{+1.4} & \le 3.1 & 1.5_{-1.0}^{+2.9} & \le 6.6 \\ 
        & Interior Only & 1.2_{-1.1}^{+0.5} & \le 2.4 & 2.6_{-2.5}^{+1.0} & \le 5.2 \\ 
        \hline
        
        Cassiopeia A & $^{44}$Ti Region & 12.0_{-4.1}^{+2.8} & 12.0_{-7.9}^{+6.4} & 4.6_{-1.6}^{+1.1} & 4.6_{-3.0}^{+2.5} \\ 
        \hline
    \end{tabular}
    \caption{Sc K$_\alpha$ fluxes and corresponding $^{44}$Ti masses for the three Milky Way SNRs. The assumed distances are the ``best'' distances in Table \ref{table:SNRs}.}
    \label{table:MW_flux}
\end{table*}

\begin{figure}[ht]
    \centering
    \includegraphics[width=\columnwidth]{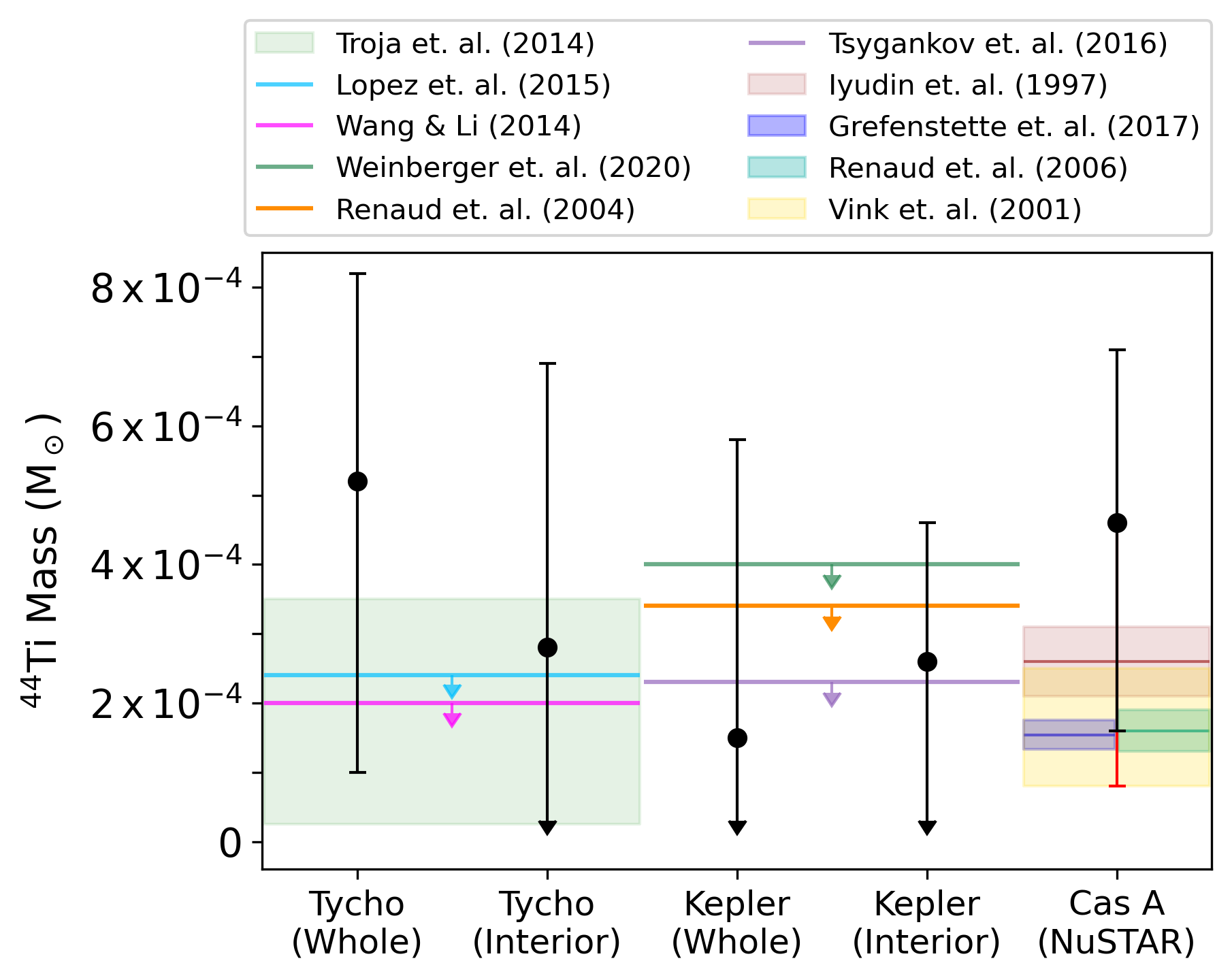}
    \caption{Inferred $^{44}$Ti masses with  {$2\sigma$} uncertainties (black) for the three Milky Way SNRs.  {The 99\% lower bound for Cas A is} shown in red. The various shaded regions and horizontal bars (with downward arrows) show the mass ranges and upper limits, respectively, of many of the works referenced in $\S$\ref{sec:results}.}
    \label{fig:massMW}
\end{figure}

\subsection{Milky Way SNRs} \label{subsec:MWresults}

In Tycho, we find an implied $^{44}$Ti mass of $(5.2_{-3.2}^{+2.4})\,\times\,10^{-4}\,\text{M}_\odot$ for the entire SNR with a $2\sigma$ lower bound of $1.0\,\times\,10^{-4}\,\text{M}_\odot$.
% 99\% lower limit is negative
When looking at the interior only, we find a mass of $(2.8_{-2.7}^{+3.5})\,\times\,10^{-4}\,\text{M}_\odot$.
\citet[Fig 4]{Troja14} found a $3\sigma$ mass range of approximately $(0.25-3.5)\,\times\,10^{-4}\,\text{M}_\odot$ at a 2.5 kpc distance,
\citet{Lopez15} found a $3\sigma$ mass upper limit of $2.4\,\times\,10^{-4}\,\text{M}_\odot$,
and \citet{Wang14} set a $3\sigma$ flux upper limit of $1.5\,\times\,10^{-5}\,\text{s}^{-1}\,\text{cm}^{-2}$ for the 68 \& 78 keV lines ---  a $^{44}$Ti mass of $2.0\,\times\,10^{-4}\,\text{M}_\odot$.
Our  mass from a Sc K$_\alpha$ gaussian agrees with these works within 2$\sigma$ as shown in Figure \ref{fig:massMW}.

In Kepler, we find a $1\sigma$ upper limit of $4.4\,\times\,10^{-4}\,\text{M}_\odot$ for the entire SNR and $3.6\,\times\,10^{-4}\,\text{M}_\odot$ for the interior.
\citet{Weinberger20} find a $2\sigma$ upper limit of $4.0\,\times\,10^{-4}\,\text{M}_\odot$ from the 68 \& 78 keV gamma-rays,
%\citet{2024MNRAS.529..999W} found a $2\sigma$ flux upper limit of $1.3\,\times\,10^{-5}\,\text{s}^{-1}\,\text{cm}^{-2}$ for the 68 keV line which implies a $^{44}$Ti mass of $4.8\,\times\,10^{-4}\,\text{M}_\odot$ using our assumptions \\
\citet{Tsygankov16} found a $3\sigma$ flux upper limit of $6.3\,\times\,10^{-6}\,\text{s}^{-1}\,\text{cm}^{-2}$ with \textit{INTEGRAL}/IBIS which implies a $^{44}$Ti mass of $2.3\,\times\,10^{-4}\,\text{M}_\odot$, and
\citet{2004ESASP.552...81R} found a $2\sigma$ flux upper limit of $1.09\,\times\,10^{-5}\,\text{s}^{-1}\,\text{cm}^{-2}$ which implies a $^{44}$Ti mass of $3.4\,\times\,10^{-4}\,\text{M}_\odot$.
The latter two masses from the cited works are calculated using our values for distance, explosion date, and mean lifetime.
The discrepancy between the upper limits of these three works and ours can likely be attributed to the large continuum flux near 4.1 keV.

In Cas A, we find an implied $^{44}$Ti mass of $(4.6_{-2.5}^{+2.0})\,\times\,10^{-4}\,\text{M}_\odot$. Our $2\sigma$ and 99\% lower limits are $1.6\,\times\,10^{-4}\,\text{M}_\odot$ and $0.8\,\times\,10^{-4}\,\text{M}_\odot$, respectively, which would be consistent with all the works mentioned below.
Once again, one can attribute the large best-fit mass to the high continuum flux near 4 keV possibly masking the true flux of the Sc K$_\alpha$ line.
This effect was especially prominent when we did a quick analysis with the entire Cas A remnant and found a somewhat improbable best-fit mass of $1.1\,\times\,10^{-3}\,\text{M}_\odot$ given the gamma-ray detections.
\citet{Iyudin97}, improving upon their initial 1994 detection, found an 1157 keV line flux of $(4.8\pm0.9)\,\times\,10^{-5}\,\text{s}^{-1}\,\text{cm}^{-2}$; using our values as mentioned above, one finds an implied $^{44}$Ti mass of $(2.6\pm0.5)\,\times\,10^{-4}\,\text{M}_\odot$. % 0.99 photons/decay
Other notable works include
\citet{Weinberger20} who found a mass of $(2.6\pm0.6)\,\times\,10^{-4}\,\text{M}_\odot$ with \textit{INTEGRAL}/SPI,
\citet{Grefenstette17} who found a mass of ($1.54\pm0.21)\,\times\,10^{-4}\,\text{M}_\odot$,
\citet{Renaud06} who measured a mass of $(1.6_{-0.3}^{+0.6})\,\times\,10^{-4}\,\text{M}_\odot$,
and \citet{Vink01} who found a mass range of $(0.8-2.5)\,\times\,10^{-4}\,\text{M}_\odot$.
Figure \ref{fig:massMW} shows the mass that we find agrees with the masses of all these works within $2\sigma$.

% {\citet{2019MNRAS.485.3288I} suggest that dust formation in the remnant, and possibly in the presupernova ejecta, could have attenuated 68 and 78 keV gamma rays.
%While there is ample evidence for dust in Cas A, there is no evidence it is substantially absorbing the SNR X-rays.
%For small regions on many lines of sight through the SNR, including the unshocked central ejecta, we and many others have fit X-ray spectra with models that include absorption.
%There is evidence for absorption at low energies (not shown here), which is consistent with the expected interstellar medium attenuation along the line of sight, but there is no evidence for any significant absorption at 4 keV (and therefore not at 68 keV either).
%As for $^{44}$Ti in large grains, which they also mention, $^{44}$Ti is found in the centers of SiC X-grains, but these are typically 0.1 microns in diameter.
%Even if in relative ``boulder'' of 10 microns, the X-ray and gamma-ray absorption by the grain is negligible.
%It is much more likely that 1.157 MeV line fluxes are different from 68 keV because of systematic effects in the measurements of the former.}

\begin{table}[tb]
    \normalsize
    \centering
    \begin{tabular}{cCC}
        \hline \hline
        Observation & \text{Line Flux} & \text{Implied Mass} \\[-1mm]
        \& Date & ($10^{-6}\,\text{s}^{-1}\,\text{cm}^{-2}$) & ($10^{-4}\,\text{M}_\odot$) \\
        \hline
         010 (Sep 2000) & 0.0_{-0.6}^{+0.8} & \le 1.7 \\
         008 (Apr 2001) & 0.2_{-0.7}^{+0.7} & 0.5_{-0.5}^{+1.4} \\
         014 (May 2003) & 0.5_{-0.3}^{+0.2} & 1.1_{-0.6}^{+0.5} \\
         040 (Jan 2007) & 0.3_{-0.5}^{+0.4} & 0.6_{-0.6}^{+1.0} \\ 
         050 (Jan 2008) & -0.1_{-0.6}^{+0.7} & \le 1.4 \\
         055 (Jan 2009) & 0.0_{-0.5}^{+0.5} & \le 1.1 \\
         060 (Dec 2009) & 0.4_{-0.7}^{+0.4} & 0.9_{-0.9}^{+0.9} \\
         065 (Dec 2010) & 0.5_{-0.6}^{+0.7} & 1.2_{-1.2}^{+1.7} \\
         067 (Dec 2011) & 0.2_{-0.5}^{+0.2} & 0.5_{-0.5}^{+0.5} \\
         069 (Dec 2012) & -0.1_{-0.4}^{+0.9} & \le 1.9 \\
         074 (Nov 2014) & 0.4_{-0.8}^{+0.6} & 1.0_{-1.0}^{+1.4} \\
         076 (Nov 2015) & 0.6_{-0.4}^{+0.5} & 1.4_{-1.0}^{+1.2} \\
         078 (Nov 2016) & 0.2_{-1.2}^{+0.6} & 0.5_{-0.5}^{+1.6} \\
         080 (Oct 2017) & 1.5_{-1.0}^{+1.3} & 3.8_{-2.4}^{+3.5} \\
         085 (Sep 2019) & 0.8_{-1.2}^{+1.3} & 2.1_{-2.1}^{+3.5} \\ 
         083 (Nov 2019) & -0.8_{-1.5}^{+1.7} & \le 2.2 \\
         086 (Nov 2020) & 1.4_{-1.2}^{+0.7} & 3.7_{-3.3}^{+1.8} \\
         088 (Dec 2021) & -0.3_{-0.5}^{+0.9} & \le 1.7 \\
         090 (Feb 2023) & 0.4_{-1.0}^{+1.0} & 1.0_{-1.0}^{+2.8} \\[1mm]
        \hline
    \end{tabular}
    \caption{Sc K$_\alpha$ fluxes with 1$\sigma$ uncertainties for the SN 1987A observations and corresponding $^{44}$Ti masses if the fluxes were unattenuated.}
    \label{table:87A_flux}
\end{table}

% SN 1987A
\subsection{Attenuation in SN 1987A} \label{subsec:87Aresults}

\begin{figure*}[ht]
    \centering
    \includegraphics[width=\textwidth]{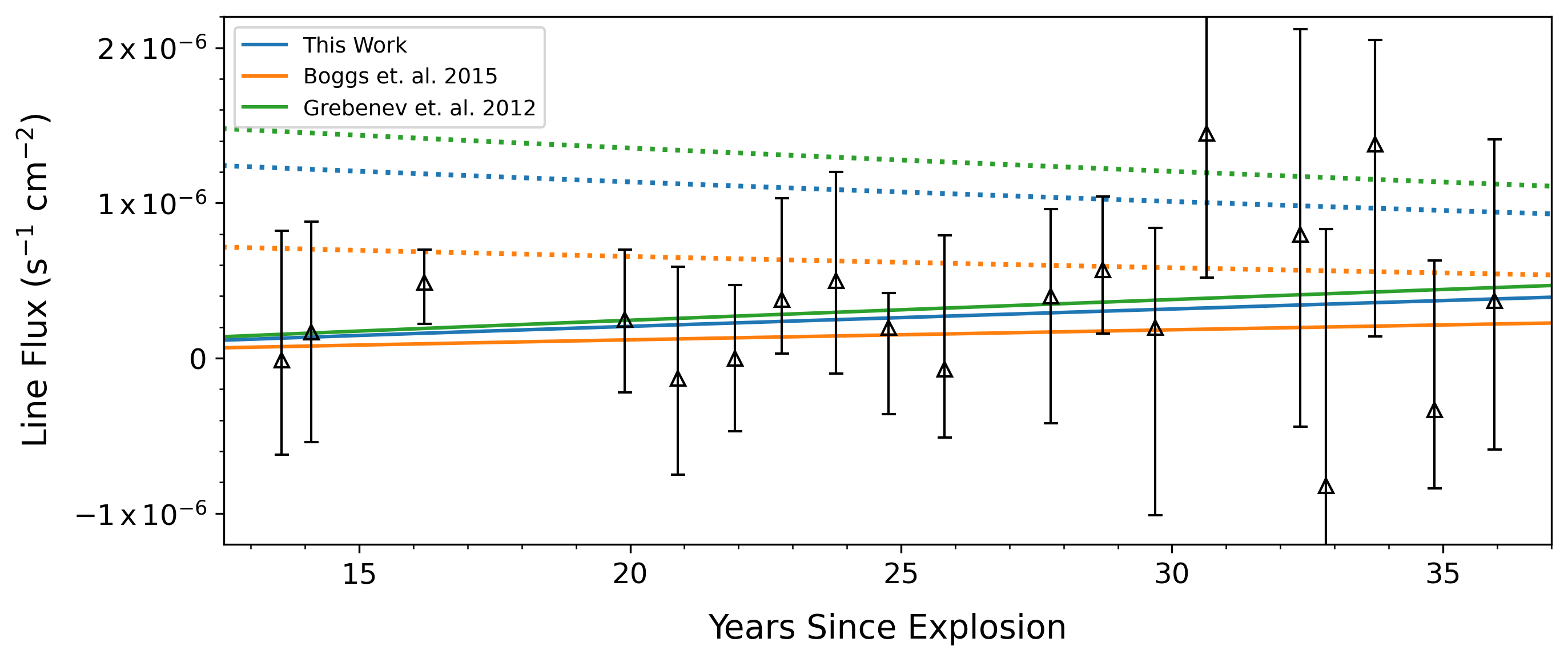}
    \caption{The fitted Sc K$_\alpha$ fluxes (black) with 1$\sigma$ uncertainties from 23 years of SN 1987A EPIC-pn observations. The solid curves show the escaping flux from model  {10HMM \citep{1988ApJ...329..820P}} with the $^{44}$Ti mass scaled to that of  {our best fit mass, the \citet{Boggs2015} best fit mass, and the \citet{Grebenev12} best fit mass. The dotted curves represent these fluxes if they were unattenuated.}}
    \label{fig:lightcurve}
\end{figure*}

We do not detect significant Sc K$_\alpha$ emission from SN 1987A  {in individual observations, even at late times, though the mean of the later points is above zero. 
Our measurements are consistent with the gamma-ray measurements, if the $^{44}$Ti remains in relatively dense (i.e., slow) regions at the times of the \textit{XMM-Newton} observations.
In the range of expected optical depths, the gamma-rays escape (opacity in Fe is $0.88\,\text{cm}^2\,\text{g}^{-1}$ at 68 keV) while the X-rays ($240\,\text{cm}^2\,\text{g}^{-1}$ at 4.1 keV) are still absorbed.}
%%% This is probably interesting enough to make it it's own paper.

{We use SN 1987A model 10HMM \citep{1988ApJ...329..820P} to show the effects of ejecta attenuation.}
This is a spherically symmetric model with a radial mixing kernel designed to move a few percent of the central $^{56}$Ni out to radial velocity $\sim\,$3000 km s$^{-1}$ to fit the early optical light curve, early emergence of $^{56}$Co gamma-ray lines, and their Compton scattered continua.
The details of the abundance distributions in the inner ejecta are not well constrained by observations, but in this model, 98\% of the $^{44}$Ti is inside 2000 km s$^{-1}$.
{Using the increasing 4 keV escape calculated for model 10HMM, we fit the line flux light curve with the total $^{44}$Ti mass as a free parameter. The relative distribution of $^{44}$Ti is unchanged. We find M($^{44}$Ti) = $(2.6 \pm 0.6)\,\times\,10^{-4}\,\text{M}_\odot$, with $\chi^2 = 26.5$ for 18 degrees of freedom. If we omit the final six points, where thermal emission increasingly interferes with the measurement of the Sc line, the result is essentially unchanged, but the fit is better ($\chi^2_\nu = 1.0$). This mass is intermediate to, and consistent with, the two gamma-ray line measurements, but we emphasize that the value is model-dependent. If we average all spectra together, or only the later spectra, we still do not see visual evidence of the Sc line, as we do in Figure \ref{fig:gauss-compare}. If the mass is ultimately very well measured, our results can be used to determine how deeply within the slowest core ejecta the $^{44}$Ti is buried.}

In Figure \ref{fig:lightcurve}, we illustrate the Sc line escape with that calculated for model 10HMM with this mass, the mass found by \citet{Boggs2015} of $(1.5\pm0.3)\,\times\,10^{-4}\,\text{M}_\odot$, and the mass found by \citet{Grebenev12} of $(3.1\pm0.8)\,\times\,10^{-4}\,\text{M}_\odot$.
% \citet{Seitenzahl14} of $(0.55\pm0.17)\,\times\,10^{-4}\,\text{M}_\odot$.

Of course, the ejecta continue to expand and thin, and the full $^{44}$Ti-decay Sc emission, and therefore its location, will be revealed in time.
We do not know exactly what to expect in the real, aspherical remnant, but the bulk of the $^{44}$Ti appears to show redshift \citep{Boggs2015} and the ejecta are elongated \citep{Arendt2023arXiv}.

\section{Conclusion} \label{sec:conclusion}

Our search for Sc K$_\alpha$ line emission produced by $^{44}$Ti decay in four SNRs with \textit{XMM-Newton} EPIC-pn data yields weak evidence, even in the cases where gamma-ray line emission has been clearly detected.
We report these as finite lower limits on the $^{44}$Ti masses, although these are unlikely to be considered detections without previously detected gamma rays.
We see a weak signal in Tycho's SNR and report a $2\sigma$ lower limit of $1.0\,\times\,10^{-4}\,\text{M}_\odot$ for the $^{44}$Ti mass for the full object.
We find marginal evidence for $^{44}$Ti decay in Kepler's SNR, which is still consistent with gamma-ray upper limits \citep[e.g.,][]{2004ESASP.552...81R}.
We find only a moderate signal in Cassiopeia A and report a 99\% lower limit of $0.8\,\times\,10^{-4}\,\text{M}_\odot$ for its $^{44}$Ti mass.
{We also find  evidence for $^{44}$Ti decay in SN 1987A, which for reasonable attenuation by the ejecta is consistent with gamma-ray measurements \citep[e.g.,][]{Boggs2015}.}
Even though the line does not stand out above the continua, even rather detailed spectral models do not account for all of the photons near 4.1 keV, so adding a line improves the fits.  

Type Ia models such as helium-shell detonation \citep[e.g.,][]{2011ApJ_Waldman} and double-degenerate explosions \citep[e.g.,][]{2022MNRAS_Pakmor} can produce $^{44}$Ti at a level $>\,10^{-3}\,\text{M}_\odot$.
We just exclude these explosion types in Tycho and Kepler as our $2\sigma$ upper limits are $8.2\,\times\,10^{-4}\,\text{M}_\odot$ and $6.6\,\times\,10^{-4}\,\text{M}_\odot$, respectively. These masses are just consistent with other double-detonation models \citep{2020ApJ...888...80L}.

In X-rays, the challenge is to identify the line photons in these bright, centuries old remnants against the large number of shock-produced line and continuum X-rays that swamp the weak radioactive decay line, whose equivalent width is tiny.
Better energy resolution, such as that of the \textit{XRISM} mission's Resolve instrument \citep{2022SPIE12181E..1SI}, might improve such measurements to the point of clear detections. 

In the young remnant of SN 1987A, the thermal emission from shocked gas is only beginning to rise, but the ejecta surrounding the $^{44}$Ti, presumably the slowest expanding matter ejected, is still somewhat optically thick at 4 keV.
This optical depth is decreasing with time since explosion, as t$^{-2}$, so these X-rays could escape at presently detectable levels before the $^{44}$Ti decays away.
Our results indicate that the bulk of the $^{44}$Ti ejected by SN 1987A was indeed among the slowest moving ejecta, at velocities below roughly 2000 km s$^{-1}$.
It is not clear how to reconcile this fact with the relatively large redshift of the $^{44}$Ti detected by \textit{NuSTAR} \citep{Boggs2015}.

\section{Acknowledgments} \label{ack}
This research has made use of data, software, and web tools obtained from the High Energy Astrophysics Science Archive Research Center (HEASARC), a service of the Astrophysics Science Division at NASA/GSFC and of the Smithsonian Astrophysical Observatory's High Energy Astrophysics Division.  {We thank the anonymous referee for helpful suggestions.}

% not sure if this is needed
\vspace{10mm}
\facilities{XMM-Newton (EPIC)}

\software{SAOImageDS9 \citep{ds9},
          matplotlib \citep{matplotlib},
          XMM-Newton SAS \citep{Gabriel04},
          XSPEC}

\clearpage
\appendix
\vspace{-5mm}
% --- can't get these tables and appendices to order themselves properly, would appreciate help from the editor ---
\section{XMM-Newton Observations Used} \label{append:obs}
\begin{deluxetable*}{ccccc}[hb]
    \tablehead{\colhead{Object} & \colhead{Obs. ID} & \colhead{PI Name} & \colhead{Obs. Date} & \colhead{Exposure Time (ks)}}
    \tablecaption{Milky Way SNR observations\label{table:obsMW}}
    \startdata
                 & 0096210101 & B. Aschenbach   & 2000 Jun 29 & 53.1 total / 9.3 filtered \\
                 & 0310590101 & F. Jansen       & 2005 Jul 03 & 33.4 / 13.5 \\
                 & 0310590201 & F. Jansen       & 2005 Aug 05 & 31.9 / 11.2 \\
                 & 0412380101 & A. Parmar       & 2006 Jul 28 & 31.9 / 15.7 \\
                 & 0412380201 & A. Parmar       & 2007 Aug 21 & 36.9 / 25.9 \\
    Tycho's SNR  & 0412380301 & A. Parmar       & 2008 Aug 08 & 46.5 / 30.2 \\
                 & 0412380401 & A. Parmar       & 2009 Aug 14 & 34.3 / 29.2 \\
                 & 0511180101 & A. Parmar       & 2007 Dec 30 & 29.1 / 13.1 \\
                 & 0801840201 & B. Williams     & 2017 Aug 04 & 42.4 / 27.4 \\
                 & 0801840301 & B. Williams     & 2017 Aug 06 & 41.9 / 29.3 \\
                 & 0801840401 & B. Williams     & 2017 Aug 10 & 37.2 / 29.0 \\
                 & 0801840501 & B. Williams     & 2017 Aug 12 & 35.9 / 25.1 \\
    \hline
                 & 0084100101 & A. Decourchelle & 2001 Mar 10 & 33.5 / 26.5 \\
    Kepler's SNR & 0842550101 & T. Sato         & 2020 Mar 19 & 140.5 / 102.1 \\
                 & 0853790101 & P. Kretschmar   & 2020 Mar 11 & 38.7 / 27.1 \\
    \hline
                 & 0097610801 & A. Brinkman     & 2000 Jul 27 & 46.7 / 10.8 \\
                 & 0110010201 & J. Bleeker      & 2000 Dec 27 & 23.4 / 5.8 \\
                 & 0110010301 & J. Bleeker      & 2000 Dec 29 & 21.4 / 4.7 \\
                 & 0110011001 & J. Bleeker      & 2001 Aug 14 & 24.0 / 5.1 \\
                 & 0110011801 & J. Bleeker      & 2001 Aug 14 & 29.7 / 5.9 \\
    Cassiopeia A & 0137550301 & F. Jansen       & 2001 Aug 22 & 45.8 / 10.2 \\
                 & 0165560101 & F. Jansen       & 2004 Jul 03 & 39.8 / 6.3 \\
                 & 0400210101 & J. Vink         & 2006 Jun 18 & 75.9 / 16.2 \\
                 & 0412180101 & A. Parmar       & 2006 Jun 22 & 58.9 / 5.5 \\
                 & 0650450201 & E. Gotthelf     & 2010 Jun 25 & 117.9 / 74.2 \\
                 & 0650450301 & E. Gotthelf     & 2010 Jun 27 & 115.3 / 74.4 \\
                 & 0650450401 & E. Gotthelf     & 2010 Jun 29 & 121.9 / 61.3 \\
    \enddata
%    \tablenotetext{}{\cite{}}
\end{deluxetable*}

%\newpage
\begin{deluxetable*}{cccc}[hbt]
    \tablehead{\colhead{Obs. ID} & \colhead{PI Name} & \colhead{Obs. Date} & \colhead{Exposure Time (ks)}}
    \tablecaption{SN 1987A observations\label{table:obs87}}
    \startdata
    \hline
                 0083250101 & B. Aschenbach   & 2001 Apr 08 & 41.0 total / 18.7 filtered \\
                 0104660101 & M. Watson       & 2000 Sep 17 & 28.7 / 4.1 \\
                 0144530101 & R. McCray       & 2003 May 10 & 132.5 / 52.8 \\
                 0406840301 & F. Haberl       & 2007 Jan 17 & 111.3 / 64.0 \\
                 0506220101 & F. Haberl       & 2008 Jan 11 & 118.8 / 83.6 \\
                 0556350101 & F. Haberl       & 2009 Jan 30 & 101.9 / 68.2 \\
                 0601200101 & F. Haberl       & 2009 Dec 11 & 91.8 / 75.7 \\
                 0650420101 & F. Haberl       & 2010 Dec 12 & 65.9 / 51.2 \\
                 0671080101 & F. Haberl       & 2011 Dec 02 & 82.5 / 61.2 \\
%   SN 1987A      &
                 0690510101 & F. Haberl       & 2012 Dec 11 & 69.9 / 58.4 \\
                 0743790101 & F. Haberl       & 2014 Nov 29 & 80.7 / 53.8 \\
                 0763620101 & F. Haberl       & 2015 Nov 15 & 67.0 / 53.9 \\
                 0783250201 & F. Haberl       & 2016 Nov 02 & 75.4 / 54.9 \\
                 0804980201 & F. Haberl       & 2017 Oct 15 & 80.5 / 35.0 \\
                 0831810101 & D. Malyshev     & 2019 Nov 27 & 36.0 / 16.9 \\
                 0852980101 & N. Schartel     & 2019 Sep 06 & 18.0 / 6.6 \\
                 0862920201 & F. Haberl       & 2020 Nov 24 & 80.8 / 52.6 \\
                 0884210101 & L. Sun          & 2021 Dec 28 & 91.7 / 68.3 \\
                 0901400101 & C. Maitra       & 2023 Feb 06 & 83.8 / 40.1 \\
    \enddata
%    \tablenotetext{}{\cite{}}
\end{deluxetable*}

\clearpage
\section{SN 1987A Spectral Fits} \label{append:87A}

SN 1987A proved a bit tricky to analyze, abundance-wise, as some K$_\alpha$ lines have little to no flux (especially in early observations).
The young remnant is continuum-dominated.
At such early times, little ejecta has been shock heated, and the ionization of the metals has had no time to catch up to the electron temperatures to produce detectable K$_\alpha$ fluorescence.

The details of all nineteen fits can be found below in Table \ref{table:87A_fits}.
All errors in this table are 90\%.
{The units of the parameters are not written explicitly in the bottom half of the table for spacing reasons, but of course they the are same as in the top half.}
The column headings are abbreviated observation numbers; the subscripts $ps$ and $ga$ represent model $pshock$ and the gaussian, respectively.
Four of the fits only required one thermal component as adding a second did not improve fit quality.
To reiterate a portion of Section \ref{sec:analysis}, we do not claim the abundances in this table are the true abundances of SN 1987A.
The goal of this work was to use physically motivated models that fit the spectra well enough to the point where we can look for flux excesses.

% big table
\begin{splitdeluxetable*}{ccccccccccBccccccccccc}
    %\rotate
    \tabletypesize{\scriptsize}
    \tablehead{\\[-8mm] \colhead{Parameter} & \colhead{008} & \colhead{010} & \colhead{014} & \colhead{040} & \colhead{050} & \colhead{055} & \colhead{060} & \colhead{065} & \colhead{067} & 
    \colhead{ {Parameter}} & \colhead{069} & \colhead{074} & \colhead{076} & \colhead{078} & \colhead{080} & \colhead{083} & \colhead{085} & \colhead{086} & \colhead{088} & \colhead{090}}
    \tablecaption{SN 1987A Spectral Fit Parameters \label{table:87A_fits}}
    \startdata % $._{-.}^{+.}$
    \hline
        ${\Delta \text{E}}_{\,6,\,ps}$\,(eV) & $82_{-80}^{+98}$ & $94_{-87}^{+96}$ & $23_{-21}^{+38}$ & $\le 12$ & $\le 15$ & $41_{-39}^{+51}$ & $41_{-36}^{+71}$ & $102_{-93}^{+82}$ & $23_{-22}^{+21}$ &
         {${\Delta \text{E}}_{\,6,\,ps}$} & $212_{-158}^{+275}$ & $13_{-12}^{+15}$ & $12_{-10}^{+14}$ & $53_{-45}^{+60}$ & $44_{-33}^{+61}$ & $195_{-185}^{+303}$ & $71_{-65}^{+121}$ & $40_{-39}^{+50}$ & $77_{-51}^{+41}$ & $\le 25$ \\
        %%%
        kT$_{ps}$ (keV) & $0.6_{-0.3}^{+0.2}$ & $1.9_{-1.6}^{+3.8}$ & $6.3_{-3.2}^{+3.5}$ & $5.3_{-3.7}^{+4.1}$ & $3.1_{-0.9}^{+3.3}$ & $2.3_{-1.3}^{+2.4}$ & $2.5_{-1.0}^{+6.0}$ & $2.0_{-1.4}^{+1.1}$ & $1.1_{-0.9}^{+0.5}$ &
         {kT$_{ps}$} & $1.6_{-0.9}^{+1.0}$ & $1.6_{-1.1}^{+0.7}$ & $1.9_{-0.7}^{+1.2}$ & $1.6_{-1.3}^{+1.1}$ & $1.8_{-0.8}^{+0.9}$ & $2.2_{-1.8}^{+3.5}$ & $4.1_{-1.2}^{+3.1}$ & $2.4_{-1.6}^{+1.7}$ & $2.1_{-0.8}^{+0.4}$ & $3.0_{-0.7}^{+0.4}$\\
        %%%
        S$_{ps}$ {\,/\,S$_\odot$} & $25.5_{-23.1}^{+45.6}$ & $\le 2.6$ & $1.1_{-0.9}^{+1.4}$ & $2.8_{-2.4}^{+4.1}$ & $2.6_{-2.1}^{+3.9}$ & $2.3_{-1.2}^{+6.0}$ & $3.9_{-2.2}^{+3.1}$ & $1.0_{-0.7}^{+3.6}$ & $\le 0.8$ &
         {S$_{ps}$\,/\,S$_\odot$} & $0.7_{-0.5}^{+0.7}$ & $0.5_{-0.4}^{+0.4}$ & $0.6_{-0.5}^{+1.3}$ & $1.0_{-0.6}^{+3.8}$ & $13.1_{-10.7}^{+21.6}$ & $10.7_{-9.9}^{+15.3}$ & $2.6_{-2.2}^{+3.1}$ & $\le 0.6$ & $8.3_{-6.6}^{+9.4}$ & $1.2_{-0.5}^{+0.7}$ \\
        %%%
        Ar$_{ps}$ {\,/\,Ar$_\odot$} & $26.5_{-21.7}^{+39.8}$ & $1.6_{-1.4}^{+4.5}$ & $0.4_{-0.3}^{+1.3}$ & $\le 1.0$ & $\le 12.2$ & $0.6_{-0.4}^{+2.7}$ & $4.3_{-3.2}^{+3.4}$ & $\le 4.0$ & $\le 1.3$ & 
         {Ar$_{ps}$\,/\,Ar$_\odot$} & $1.0_{-0.9}^{+1.8}$ & $\le 0.3$ & $0.5_{-0.4}^{+0.9}$ & $8.5_{-7.6}^{+12.6}$ & $13.3_{-10.2}^{+30.5}$ & $\le 0.2$ & $5.2_{-4.4}^{+8.8}$ & $1.0_{-0.9}^{+1.5}$ & $5.5_{-4.5}^{+8.9}$ & $1.1_{-0.7}^{+1.1}$ \\
        %%%
        Ca$_{ps}$ {\,/\,Ca$_\odot$} & $31.9_{-19.7}^{+49.0}$ & $\le 2.2$ & $\le 3.8$ & $\le 2.4$ & $\le 28.4$ & $\le 9.2$ & $0.8_{-0.7}^{+1.5}$ & $\le 11.2$ & $\le 3.3$ & 
         {Ca$_{ps}$\,/\,Ca$_\odot$} & $2.7_{-2.5}^{+4.3}$ & $3.5_{-3.1}^{+7.1}$ & $2.8_{-2.4}^{+4.6}$ & $\le 6.5$ & $34.8_{-31.5}^{+61.8}$ &  $38.7_{-35.6}^{+83.5}$ & $16.5_{-12.7}^{+13.8}$ & $3.6_{-3.5}^{+5.4}$ & $13.5_{-10.9}^{+23.8}$ & $1.4_{-0.7}^{+1.9}$ \\
        %%%
        Fe$_{ps}$ {\,/\,Fe$_\odot$} & $1.2_{-1.1}^{+3.7}$ & $1.2_{-1.0}^{+6.3}$ & $\le 1.1$ & $1.3_{-1.2}^{+2.0}$ & $1.6_{-1.5}^{+1.5}$ & $1.5_{-1.3}^{+2.4}$ & $\le 0.1$ & $1.5_{-1.3}^{+8.7}$ & $1.9_{-1.6}^{+3.6}$ &
         {Fe$_{ps}$\,/\,Fe$_\odot$} & $0.5_{-0.4}^{+0.8}$ & $0.4_{-0.3}^{+0.6}$ & $\le 0.1$ & $2.8_{-2.6}^{+8.2}$ & $4.2_{-4.1}^{+9.4}$ & $\le 0.1$ & $\le 1.7$ & $\le 1.0$ & $5.9_{-5.1}^{+10.3}$ & $0.5_{-0.2}^{+0.3}$ \\
        %%%
        $\tau_{ps}$ ($10^{10}\,\text{cm}^{-3}\,s$) & $3.70_{-2.44}^{+5.30}$ & $2.87_{-0.89}^{+4.41}$ & $5.85_{-5.24}^{+6.98}$ & $6.40_{-5.02}^{+9.00}$ & $3.44_{-0.49}^{+0.60}$ & $7.57_{-3.48}^{+30.8}$ & $4.17_{-2.10}^{+4.72}$ & $3.52_{-2.35}^{+8.67}$ & $13.6_{-6.1}^{+19.8}$ &
         {$\tau_{ps}$} & $2.78_{-1.68}^{+5.55}$ & $2.45_{-2.16}^{+4.60}$ & $3.55_{-3.35}^{+7.65}$ & $4.58_{-3.67}^{+9.35}$ & $28.1_{-9.9}^{+38.3}$ & $0.20_{-0.18}^{+0.28}$ & $0.52_{-0.50}^{+0.63}$ & $3.46_{-3.22}^{+4.98}$ & $74.7_{-19.0}^{+46.0}$ & $95.0_{-34.3}^{+97.2}$ \\
        %%%
        z$_{ps}$ ($10^{-3}$) & $-6.4_{-14.0}^{+10.8}$ & $19.7_{-25.7}^{+21.7}$ & $-11.7_{-29.2}^{+17.2}$ & $7.1_{-20.4}^{+16.9}$ & $6.4_{-16.5}^{+45.8}$ & $1.5_{-7.2}^{+4.5}$ & $-1.9_{-6.4}^{+6.5}$ & $9.6_{-27.1}^{+36.3}$ & $22.2_{-92.7}^{+70.1}$ &
         {z$_{ps}$} & $24.3_{-28.1}^{+30.8}$ & $0.3_{-0.6}^{+1.1}$ & $13.9_{-20.7}^{+19.9}$ & $17.9_{-23.5}^{+58.6}$ & $-2.1_{-2.1}^{+2.6}$ & $-21.6_{-28.5}^{+45.2}$ & $-25.8_{-18.3}^{+37.0}$ & $31.4_{-44.4}^{+39.1}$ & $6.5_{-5.0}^{+4.1}$ & $2.7_{-4.5}^{+2.5}$\\
        %%%
        norm$_{ps}$ ($10^{-3}$) & $0.02_{-0.01}^{+0.01}$ & $0.11_{-0.10}^{+0.08}$ & $0.08_{-0.06}^{+0.16}$ & $0.35_{-0.29}^{+0.56}$ & $0.25_{-0.15}^{+0.94}$ & $0.49_{-0.32}^{+0.59}$ & $0.37_{-0.21}^{+0.42}$ & $1.23_{-0.96}^{+1.23}$ & $2.67_{-2.40}^{+7.15}$ & 
         {norm$_{ps}$} & $3.30_{-2.38}^{+1.79}$ & $3.00_{-2.31}^{+2.86}$ & $3.04_{-2.66}^{+2.26}$ & $3.12_{-2.63}^{+3.39}$ & $1.06_{-0.90}^{+0.21}$ & $1.47_{-1.36}^{+2.35}$ & $2.59_{-1.11}^{+1.64}$ & $1.89_{-1.65}^{+2.74}$ & $1.72_{-1.69}^{+1.20}$ & $3.88_{-1.80}^{+1.12}$ \\[1mm]
        \hline
        ${\Delta \text{E}}_{\,6,\,nei}$\,(eV) & $-$ & $-$ & $\le 10$ & $61_{-31}^{+41}$ & $104_{-46}^{+96}$ & $162_{-82}^{+160}$ & $132_{-113}^{+281}$ & $\le 13$ & $51_{-47}^{+60}$ &
         {${\Delta \text{E}}_{\,6,\,nei}$} & $73_{-69}^{+93}$ &
        $52_{-46}^{+52}$ & $37_{-32}^{+52}$ & $18_{-13}^{+27}$ & $-$ & $140_{-120}^{+140}$ & $97_{-88}^{+78}$ & $50_{-45}^{+59}$ & $-$ & $\le 19$ \\
        %%%
        kT$_{nei}$ (keV) & $-$ & $-$ & $4.5_{-3.5}^{+7.3}$ & $5.4_{-5.0}^{+6.2}$ & $3.3_{-3.1}^{+5.4}$ & $2.6_{-2.4}^{+9.0}$ & $4.4_{-2.0}^{+8.5}$ & $1.5_{-0.4}^{+1.0}$ & $4.0_{-2.4}^{+4.2}$ &
         {kT$_{nei}$} & $2.0_{-0.7}^{+1.6}$ & $3.0_{-1.0}^{+2.3}$ & $1.8_{-0.4}^{+0.7}$ & $2.3_{-0.9}^{+1.7}$ & $-$ & $1.5_{-0.7}^{+1.0}$ & $1.0_{-0.4}^{+0.4}$ & $1.9_{-0.6}^{+1.0}$ & $-$ & $1.0_{-0.2}^{+0.2}$\\
        %%%
        S$_{nei}$ {\,/\,S$_\odot$} & $-$ & $-$ & $37.6_{-35.4}^{+42.2}$ & $38.9_{-37.8}^{+48.9}$ & $7.9_{-6.2}^{+12.1}$ & $9.8_{-9.1}^{+24.3}$ & $2.3_{-1.9}^{+6.7}$ & $33.5_{-22.8}^{+44.9}$ & $38.1_{-18.9}^{+28.5}$ &
         {S$_{nei}$\,/\,S$_\odot$} & $30.5_{-15.1}^{+42.5}$ & $29.0_{-13.9}^{+17.7}$ & $28.3_{-10.6}^{+25.4}$ & $37.8_{-20.7}^{+58.6}$ & $-$ & $6.8_{-4.6}^{+11.1}$ & $8.9_{-6.9}^{+8.7}$ & $3.3_{-1.8}^{+3.7}$ & $-$ & $2.4_{-1.7}^{+2.4}$\\
        %%%
        Ar$_{nei}$ {\,/\,Ar$_\odot$} & $-$ & $-$ & $142_{-122}^{+153}$ & $42.0_{-41.3}^{+42.7}$ & $34.7_{-24.6}^{+56.5}$ & $5.1_{-4.7}^{+11.7}$ & $\le 12.4$ & $25.0_{-21.5}^{+43.7}$ & $19.5_{-14.8}^{+21.7}$ &
         {Ar$_{nei}$\,/\,Ar$_\odot$} & $21.5_{-17.6}^{+29.3}$ & $8.1_{-6.5}^{+14.5}$ & $20.5_{-10.1}^{+27.6}$ & $21.4_{-16.7}^{+49.4}$ & $-$ & $22.6_{-16.4}^{+41.4}$ & $\le 5.7$ & $2.5_{-1.9}^{+3.7}$ & $-$ & $2.2_{-1.8}^{+3.1}$\\
        %%%
        Ca$_{nei}$ {\,/\,Ca$_\odot$} & $-$ & $-$ & $186_{-165}^{+235}$ & $\le 44.9$ & $40.6_{-25.7}^{+78.3}$ & $7.1_{-6.8}^{+9.5}$ & $\le 29.8$ & $11.7_{-10.9}^{+17.6}$ & $18.3_{-14.3}^{+35.1}$ &
         {Ca$_{nei}$\,/\,Ca$_\odot$} & $9.3_{-8.5}^{+9.2}$ & $16.3_{-11.1}^{+30.3}$ & $13.4_{-11.7}^{+29.9}$ & $48.3_{-34.4}^{+66.8}$ & $-$ & $2.21_{-2.11}^{+4.47}$ & $\le 16.9$ & $2.0_{-1.8}^{+3.6}$ & $-$ & $1.5_{-0.8}^{+4.7}$ \\
        %%%
        Fe$_{nei}$ {\,/\,Fe$_\odot$} & $-$ & $-$ & $53.6_{-47.5}^{+53.7}$ & $31.1_{-30.7}^{+47.1}$ & $24.3_{-18.8}^{+64.5}$ & $16.1_{-14.0}^{+15.4}$ & $12.3_{-10.2}^{+20.4}$ & $30.4_{-22.3}^{+57.5}$ & $13.8_{-11.4}^{+11.8}$ &
         {Fe$_{nei}$\,/\,Fe$_\odot$} & $21.5_{-14.2}^{+26.0}$ & $75.9_{-42.8}^{+79.7}$ & $21.7_{-10.9}^{+16.6}$ & $19.6_{-13.5}^{+22.3}$ & $-$ & $5.1_{-4.8}^{+8.9}$ & $57.2_{-28.3}^{+51.3}$ & $5.6_{-3.5}^{+7.2}$ & $-$ & $3.4_{-3.1}^{+4.9}$ \\
        %%%
        $\tau_{nei}$ ($10^{10}\,\text{cm}^{-3}\,s$) & $-$ & $-$ & $0.08_{-0.06}^{+0.17}$ & $1.45_{-0.98}^{+3.78}$ & $1.68_{-1.50}^{+15.6}$ & $3.43_{-2.52}^{+5.50}$ & $3.71_{-3.13}^{+12.1}$ & $48.0_{-33.3}^{+92.1}$ & $10.1_{-5.4}^{+17.3}$ &
         {$\tau_{nei}$} & $42.9_{-34.5}^{+50.9}$ & $15.6_{-8.0}^{+9.5}$ & $22.2_{-11.9}^{+22.2}$ & $32.4_{-20.7}^{+95.8}$ & $-$ & $94.3_{-86.4}^{+93.4}$ & $89.6_{-26.1}^{+65.8}$ & $35.9_{-26.7}^{+54.9}$ & $-$ & $62.6_{-32.1}^{+80.4}$ \\
        %%%
        z$_{nei}$ ($10^{-3}$) & $-$ & $-$ & $-51.3_{-17.8}^{+17.2}$ & $-8.8_{-8.4}^{+9.9}$ & $-12.1_{-7.4}^{+10.5}$ & $-53.6_{-91.3}^{+88.1}$ & $-13.2_{-25.7}^{+25.5}$ & $-5.8_{-5.8}^{+9.1}$ & $0.6_{-4.8}^{+4.0}$ & 
         {z$_{nei}$} & $-7.2_{-4.2}^{+8.0}$ & $3.4_{-4.0}^{+4.8}$ & $-7.7_{-3.6}^{+6.2}$ & $-3.3_{-4.8}^{+6.1}$ & $-$ & $-0.9_{-9.4}^{+9.9}$ & $-18.3_{-13.8}^{+11.6}$ & $-0.9_{-3.2}^{+2.1}$ & $-$ & $-12.9_{-6.0}^{+13.0}$\\
        %%%
        norm$_{nei}$ ($10^{-3}$) & $-$ & $-$ & $0.03_{-0.02}^{+0.02}$ & $0.05_{-0.04}^{+0.08}$ & $0.05_{-0.04}^{+0.12}$ & $0.06_{-0.05}^{+0.16}$ & $0.05_{-0.04}^{+0.07}$ & $0.09_{-0.07}^{+0.06}$ & $0.07_{-0.04}^{+0.04}$ &
         {norm$_{nei}$} & $0.09_{-0.06}^{+0.07}$ & $0.09_{-0.05}^{+0.06}$ & $0.10_{-0.05}^{+0.05}$ & $0.10_{-0.07}^{+0.07}$ & $-$ & $0.56_{-0.36}^{+0.87}$ & $1.13_{-0.81}^{+0.75}$ & $0.74_{-0.41}^{+1.02}$ & $-$ & $2.06_{-1.08}^{+1.26}$\\[1mm]
        \hline
        $\alpha_{pow}$ & $2.62_{-0.29}^{+0.37}$ & $2.06_{-0.61}^{+0.97}$ & $3.06_{-0.74}^{+0.92}$ & $3.27_{-0.22}^{+1.76}$ & $3.35_{-0.78}^{+1.90}$ & $3.36_{-0.24}^{+0.59}$ & $3.44_{-0.13}^{+0.49}$ & $3.30_{-0.24}^{+0.55}$ & $2.91_{-0.53}^{+0.43}$ &
         {$\alpha_{pow}$} & $3.25_{-0.46}^{+1.01}$ & $2.93_{-0.19}^{+0.33}$ & $2.98_{-0.24}^{+0.62}$ & $2.90_{-0.21}^{+0.32}$ & $3.20_{-0.12}^{+0.24}$ & $3.05_{-0.48}^{+1.04}$ & $3.53_{-1.28}^{+2.14}$ & $3.29_{-0.55}^{+1.10}$ & $2.85_{-0.15}^{+0.10}$ & $3.04_{-0.54}^{+0.54}$\\
        %%%
        norm$_{pow}$ ($10^{-3}$) & $0.15_{-0.03}^{+0.04}$ & $0.09_{-0.06}^{+0.07}$ & $0.20_{-0.10}^{+0.20}$ & $1.37_{-0.50}^{+1.51}$ & $2.06_{-0.63}^{+0.23}$ & $2.47_{-0.97}^{+1.26}$ & $3.40_{-0.58}^{+1.51}$ & $2.92_{-0.95}^{+1.56}$ & $2.51_{-1.94}^{+2.17}$ &
         {norm$_{pow}$} & $2.75_{-1.59}^{+1.98}$ & $3.00_{-1.30}^{+1.96}$ & $3.05_{-1.39}^{+1.97}$ & $3.19_{-1.47}^{+2.38}$ & $5.49_{-1.17}^{+2.59}$ & $3.37_{-1.69}^{+2.03}$ & $3.54_{-3.51}^{+3.83}$ & $3.81_{-2.33}^{+2.63}$ & $3.17_{-0.80}^{+1.47}$ & $1.17_{-0.52}^{+0.88}$ \\[1mm]
        \hline
        $\text{E}_\text{\,ga}$\,(keV) & $4.22_{-0.12}^{+0.28}$ & $4.21_{-0.10}^{+0.29}$ & $4.32_{-0.24}^{+0.17}$ & $4.10$ & $4.26_{-0.12}^{+0.24}$ & $4.10$ & $4.10_{-0.01}^{+0.40}$ & $4.15_{-0.05}^{+0.35}$ & $4.33_{-0.23}^{+0.18}$ &
         {$\text{E}_\text{\,ga}$} & $4.26_{-0.16}^{+0.25}$ & $4.10_{-0.01}^{+0.41}$ & $4.27_{-0.18}^{+0.24}$ & $4.32_{-0.22}^{+0.19}$ & $4.13_{-0.03}^{+0.30}$ & $4.23_{-0.13}^{+0.27}$ & $4.12_{-0.02}^{+0.38}$ & $4.19_{-0.08}^{+0.31}$ & $4.10$ & $4.13$ \\
        %%%
        $\text{F}_\text{ga}$\,($10^{-6}\,\text{s}^{-1}\,\text{cm}^{-2}$) & $0.2_{-1.2}^{+1.2}$ & $0.0_{-0.9}^{+1.5}$ & $0.5_{-0.5}^{+0.4}$ & $0.3_{-0.8}^{+0.7}$ & $-0.1_{-0.9}^{+1.0}$ & $0.0_{-0.7}^{+0.7}$ & $0.4_{-1.0}^{+0.7}$ & $0.5_{-1.0}^{+1.2}$ & $0.2_{-0.9}^{+0.6}$ &
         {$\text{F}_\text{ga}$} & $-0.1_{-0.7}^{+1.4}$ & $0.4_{-1.4}^{+1.0}$ & $0.6_{-0.7}^{+0.9}$ & $0.2_{-1.8}^{+1.2}$ & $1.5_{-1.8}^{+2.0}$ & $-0.8_{-2.4}^{+2.7}$ & $0.8_{-2.6}^{+2.9}$ & $1.4_{-1.8}^{+1.3}$ & $-0.3_{-1.0}^{+1.4}$ & $0.4_{-1.6}^{+1.7}$\\[1mm]
        \hline
        C-Stat/dof & 105/109 & 105/134 & 94/107 & 239/219 & 304/284 & 376/324 & 334/384 & 339/329 & 306/381 &
         {C-Stat/dof} & 337/370 & 302/354 & 337/370 & 422/447 & 428/479 & 209/198 & 86/104 & 467/453 & 1153/1091 & 1548/1484 \\[1mm]
    \enddata
%    \tablenotetext{}{\cite{}}
\end{splitdeluxetable*}

%\clearpage
\section{MCMC} \label{append:mcmc}
\begin{figure*}[ht]
    \centering
    \includegraphics[width=\textwidth]{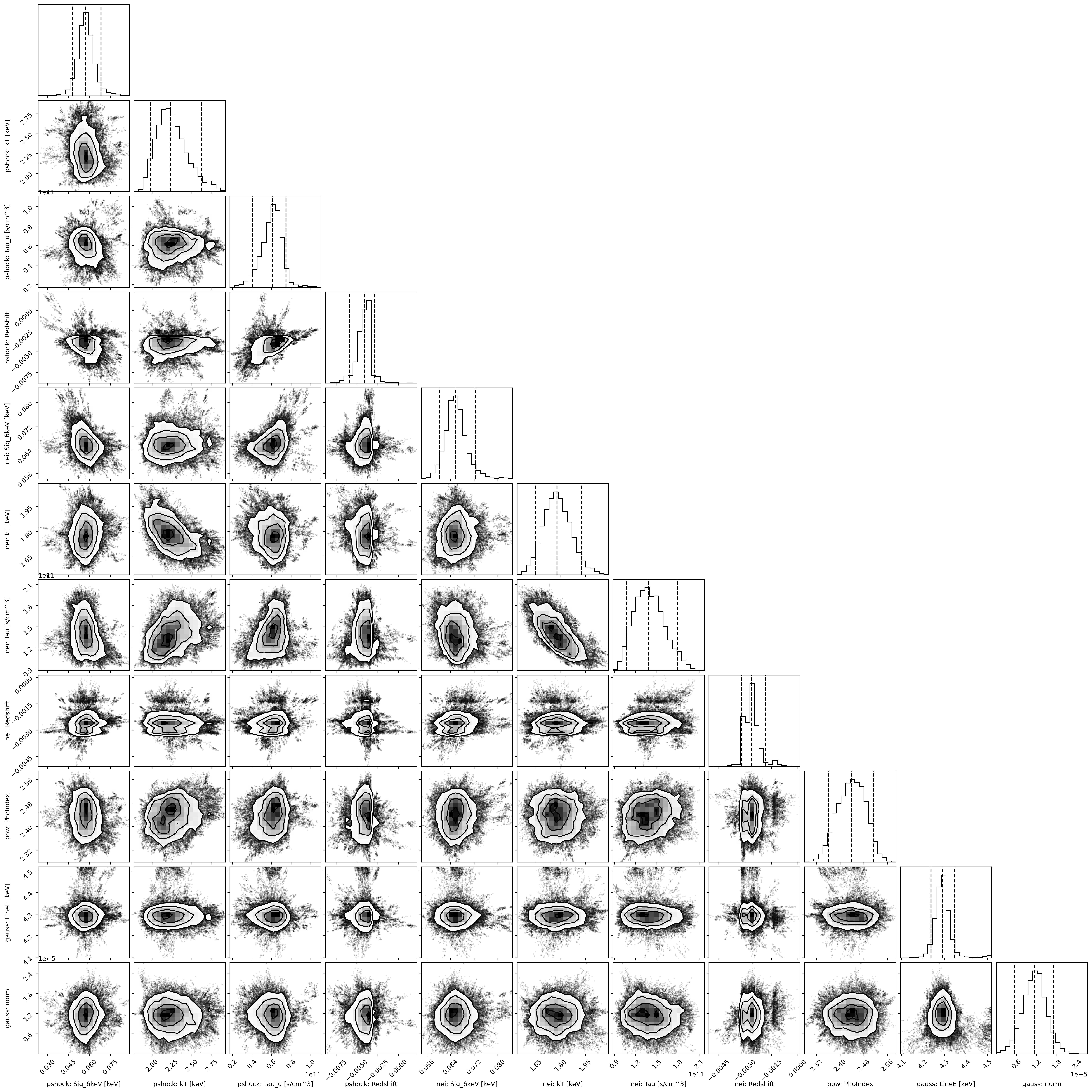}
    \caption{Corner plot for Table \ref{table:cas_a_nustar}  {(excluding the elemental abundances)}. The  {bottom} row is the gaussian flux parameter. There is little correlation between this and other parameters, which justifies using using physically constrained spectral models.}
    \label{fig:casA_mcmc}
\end{figure*}

\clearpage
\bibliography{paper}{}
\bibliographystyle{aasjournal}

\end{document}